\documentclass[conference,compsoc]{IEEEtran}
\usepackage{graphicx}
\usepackage{hyperref}
\usepackage{subfig}
\usepackage{xcolor}

\usepackage{array, multirow}

\usepackage{algorithm}
\usepackage{algpseudocode}
\newtheorem{theorem}{Theorem}

\newcommand{\FIXME}[1]{ 
}

\begin{document}
	%
	\title{Optimized Graph Based Routing Algorithm for the Angara Interconnect}

	
	\author{\IEEEauthorblockN{Anatoly Mukosey\IEEEauthorrefmark{1},
	Alexander Semenov\IEEEauthorrefmark{1},\IEEEauthorrefmark{2} and
	Alexander Tretiakov\IEEEauthorrefmark{1}}
	\IEEEauthorblockA{\IEEEauthorrefmark{1}JSC NICEVT \\Scientific and Research Centre of \\Electronic Computer Technology, Russian Federation\\
	\IEEEauthorblockA{\IEEEauthorrefmark{2}
		National Research University Higher School of Economics, Russian Federation \\
	Email: semenov@nicevt.ru}}
}


	\maketitle
	\begin{abstract}
		JSC NICEVT has developed the Angara high-speed interconnect with 4D torus topology.
		The Angara interconnect router implements deterministic routing based on the bubble flow control, a direction order routing (DOR) and direction bits rules. The router chip also supports non standard First Step	/ Last Step for bypassing failed nodes and links, these steps can violate the DOR rule. 
        In the previous work we have proposed an algorithm for generation and analysis of routing tables that guarantees no deadlocks in the Angara interconnect. It is based on a breadth-first search algorithm in a graph and it practically does not take into consideration communication channel load. Also we have never evaluated the influence of routing table generation algorithm on the performance of a real-world Angara based cluster. 
        In this paper we present a routing graph notation that provides a possibility to build routes in the torus topology of the Angara interconnect. We propose a deadlock-free routing
        algorithm based on a fast single-source shortest path algorithm for the deterministic Angara routing with a single virtual channel.
        We evaluated the considered routing algorithms 
        on a 32-node Desmos cluster system and benchmarked the proposed algorithm performance improvement of 11.1\% for the Alltoall communication pattern and of more than 5\% for the FT and IS application kernels.
	\end{abstract}
	

	%
	\IEEEpeerreviewmaketitle

\section{Introduction and Motivation}
High performance computing (HPC) systems are costly hardware and optimization of application performance is a very challenging problem. The critical component of HPC system is an interconnect that properties stand
behind the scalability of any MPI based parallel algorithm. The interconnect is largely controlled by routing algorithms which determine how to
move packets through an interconnect topology. 

We consider the Angara interconnect~\cite{simonov2014pervoye} that was developed at JSC NICEVT in Russia. Angara is the low-latency, high bandwidth interconnect with 4D torus topology, the minimum obtained MPI-latency between 2 adjacent nodes is 850 ns. The Angara-C1 \cite{agarkov2016performance} and Desmos \cite{stegailov2019angara} cluster systems are based on the Angara interconnect. Many authors have obtained results with use of the Angara interconnect \cite{khalilov2018optimization,ostroumova2019reactive,polyakov2018,stegailovvasp2019,tolstykh2018structure,stegailov2020}. Now the maximum node number of Angara based high performance computing systems is 96 nodes.

Deterministic routing in the Angara interconnect is based on the direction order routing (DOR)~\cite{adiga2005blue,scott1996cray} implemented using direction bits \cite{scott1996cray} and additional First and Last Steps~\cite{scott1996cray} for bypassing failed nodes and links. Implementation of the First and Last Steps method allows to violate the DOR rule. In the previous work \cite{mukosey2019extended} we have proposed an algorithm for generation and analysis of routing tables that guarantees no deadlocks in the Angara interconnect. The proposed algorithm increases the number of different routable systems and improves fault tolerance, the algorithm is based on a breadth-first search algorithm in a graph and practically does not take into consideration a communication channel load. Also we have never evaluated the influence of a routing table generation algorithm on the performance of a real-world Angara based cluster. 

In this work we focus on development of an advanced routing table generation algorithm and on the routing algorithm evaluation on the Desmos Angara based cluster. 

\subsection{Related Work}

Routing algorithms can be divided in two categories: oblivious and
adaptive algorithms \cite{dally2004principles}. An oblivious algorithm 
chooses a route for each source-destination pair $(s, d)$ without considering any information about the network's traffic. An adaptive algorithm adapts to the current network traffic conditions. Deterministic routing algorithms is a subset of oblivious algorithms and choose the same path between $s$ and $d$,
even if there are multiple possible paths. 

The network property path diversity is a routing possibility to provide multiple minimal routes between most pairs of nodes, it adds to the robustness of the network by balancing load across channels and allowing the network to tolerate faulty channels and nodes. Balancing the load across the communication channels affects the network performance. 

The \emph{edge-forwarding index} is the maximum channel load $\pi = \max \gamma_c$ for each communication channel $c$ from a set $C$ of all communication channels for a network \cite{heydemann1989forwarding}. The \emph{channel load} $\gamma_c$ is a number of routes crossing network channel $c$. The edge-forwarding index addresses link contention and thus defines the balancedness and the quality of the routing table. 

The \emph{perfect channel load} is an average channel load $\gamma_{perfect} = \frac{\sum \gamma_c}{\mid C \mid}$ in a system with a routing table, in which each route has minimal length. By analogy with \cite{sancho2003routing} we define the  \emph{deviation} metric $\sigma(k,\mathcal{R}) = \sqrt[k]{\frac{1}{\mid C \mid}\sum\limits_{c \in C}{|\gamma_{perfect} - \gamma_c}\mid^k}$ for a routing table $R$. The deviation metric characterizes a deviation of channel loads from a perfect channel load and can be more descriptive than the edge-forwarding index.

\FIXME{Bisection, why we use only max li?}

Optimal deterministic
routing (with a minimal edge-forwarding index) in arbitrary
networks is NP-hard \cite{saad1993complexity}. 
Several approaches and techniques exist to resolve balancing problem in general and torus topology networks.

The first approach to implement load balanced routing for torus networks over dimension (or direction) order routing is to forward packets depending on relative position of a source-destination network node pair \cite{montanana2009balanced}. This algorithm is simple, it can be implemented in hardware, but it does not take into consideration nonsymmectric topology cases and channel faults.

\FIXME{BlueGene?}

Instead of choosing routing paths locally at each router it is possible to use global routing, all communications can be optimized for each source-destination node pairs, particularly in case of communication pattern presence. 
Mixed integer linear programming (MILP) can be used to obtain an optimal solution of the global routing problem \cite{kinsy2009application,abdel2011transcom} for arbitrary topology, but because of the computational complexity, this approach does not scale well. 

In \cite{abdel2016scalable} A. Abdel-Gawad has proposed 
a decoupled approach SGR that consists of discovering the best possible channel loads and constructing source-destination pairs that achieve
such optimal channel loads. The best channel loads are obtained by cast the global routing problem to a linear programming (LP) problem, which is solvable in polynomial time. But the obtained best channel loads imply a source-destination flow splitting and the second SGR stage of constructing source-destination pair paths is based on the greedy algorithm that achives optimal channel loads through a possibility of additional paths for a source-destination pair. MILP and LP formulations take into consideration injection flows, as a result the global routing can be application specific.

Graph methods are natural way to perform balancing of routing paths. A network can be modeled as a directed \emph{network graph} $G(N,C)$ in which $N$ represents the set of network nodes and $C$ represents the
set of communication links between the nodes. The \emph{channel dependency graph} $CDG = G(C,E)$ is a directed graph with the edge set $C$ of a network as vertices, the routing function $R$ defines edges $e = (c_i, c_j)$ as a possible path step in a network. \cite{dally1988deadlock,schwiebert2001deadlock} states that a routing function is deadlock-free if the corresponding channel dependency graph is acyclic.

T. Hoefler et al. have proposed \cite{hoefler2009optimized} a balancing routing algoritm for arbitrary interconnect topologies based on the single-source shortest path
(SSSP) Dijkstra algorithm \cite{dijkstra1959note}. For each network node 
routing paths are constructed by the single-source shortest paths algorithm through the network graph. The balancing is reached through an
edge weight, which is incremented by one for all edges of a path from a source to a destination node. A $P$-$SSSP$ algorithm variant for each source node builds routing paths to all possible destination nodes, and then modifies edge weights. $P$-$SSSP$ does not balance the edge-forwarding index ideally, but its runtime includes $n$ SSSP calls, where $n$ is a number of network nodes. A more accurate $P^2$-$SSSP$ algorithm performs SSSP for each source-destination pair and updates the edge weights every time. Clearly, the $P^2$-$SSSP$ runtime is dominated by $n^2$ calls of SSSP. The SSSP based algorithms might introduce cyclic dependencies, which might lead to network deadlocks. For that reason the second part of an algorithm called DFSSSP \cite{domke2011deadlock} places all obtained paths to a channel dependency graph and then breaks cycles in the graph by moving paths to virtual graph layers, each layer is assigned to a virtual channel. 

The breadth-first and depth-first search algorithms through the network graph can be used to generate routing tables for arbitrary irregular network topology \cite{sancho2004effective}.


A genetic algorithm is a well-known search heuristic to solve NP-hard problems which relies on the process of natural selection. But for large networks the problem is the time complexity of genetic algorithms.

In \cite{sancho2000improving} a heuristic iterative load balancing algoritms is proposed. For all source-destination node pairs all possible paths are constructed, and then in each step the channel with highest load value is determined, a routing path crossing this channel is selected to be removed and excluded from the consideration if there is more than one routing path between a source and a destination nodes of this routing path. The algorithm finishes when the number of routing paths between every pair of nodes is reduced down to the unit.




The major contributions of our work are:
\begin{itemize}
\item We present full description of a routing graph notation that provides a possibility to build routes in the Angara interconnect with a routing that places restrictions on a further packet route with the previous route steps in mind.
\item We propose a deadlock-free routing
algorithm based on a fast SSSP algorithm for the deterministic routing of the Angara interconnect with a single virtual channel.
\item We evaluate the SSSP based proposed algorithm, a genetic algorithm and a previously proposed BFS based algorithm and compare the edge-forwarding index, routing time and application performance for synthetic communication patterns and NPB benchmarks on the Desmos cluster.
\end{itemize}

The paper is structured as follows. Section 2 descibes the Angara interconnect architecture and its routing. 
In Section 3 we present the routing graph notation and a condition and an algorithm for edges adding to a routing graph that does not lead to deadlocks. Section 4 presents the previously proposed BFS based algorithm, devised genetic algorithm and proposed SSSP based algorithm. In Section 5 we discuss the improvements of
the SSSP routing for balancedness quality, runtime and application performance. The last Section 6 concludes the paper.
	
\section{The Angara Interconnect Architecture}	

The Angara interconnect is a Russian-designed communication
network with 4D torus topology. The interconnect ASIC was developed by JSC NICEVT and manufactured by TSMC with the 65-nm process.

\begin{figure}[h]
	\begin{center}
		\includegraphics[height=7cm]{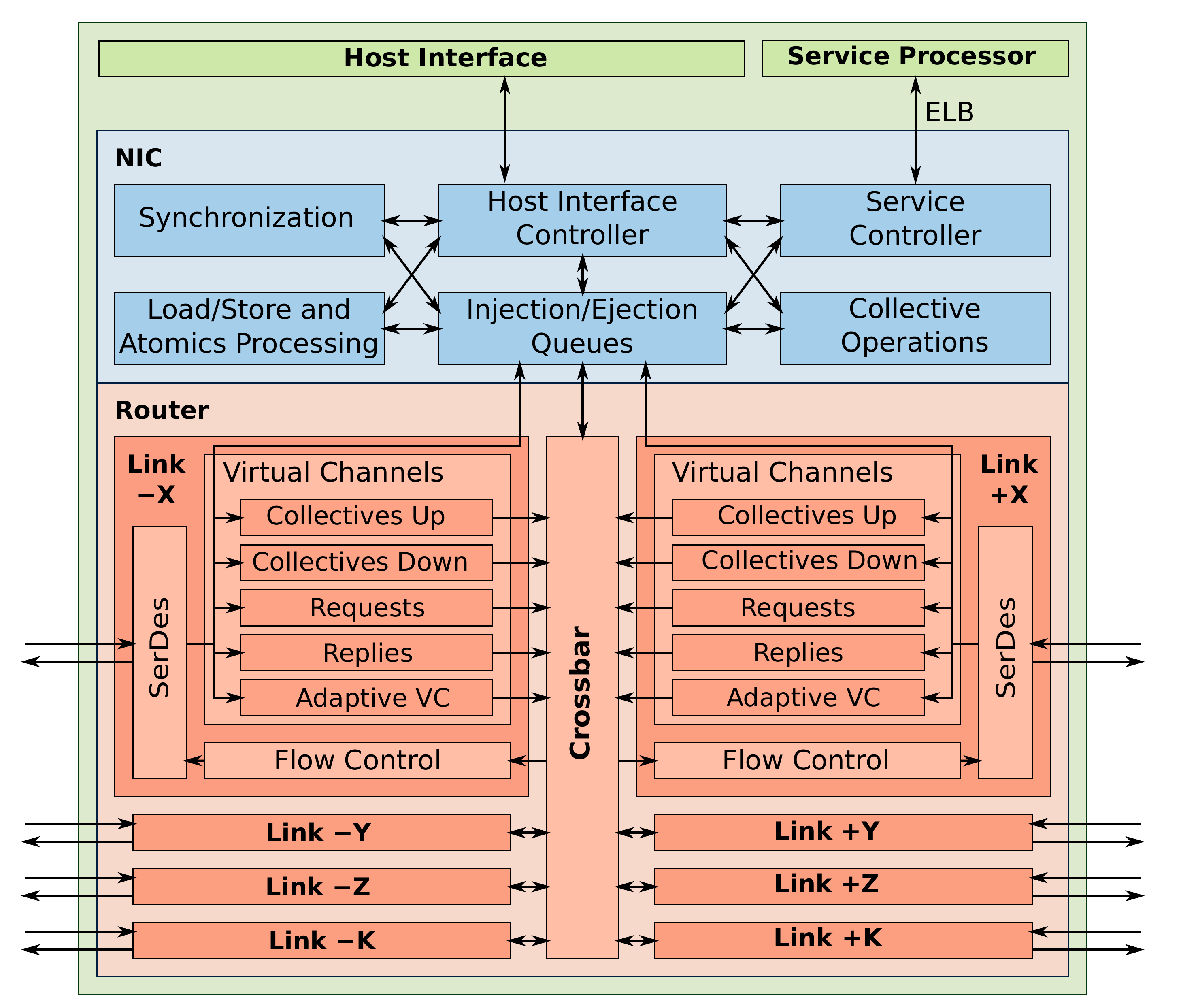}
		\caption{The Angara ASIC architecture.}
		\label{fig:asic}
	\end{center}
\end{figure}

The Angara ASIC chip (Figure \ref{fig:asic}) consists of an adapter and a router. The router contains 8 link blocks that are connected via a crossbar. The crossbar can simultaneously transmit flits (128 bits) from links if there are no conflicts.

In each direction there are five
virtual channels: two channels for deterministic
routing (blocking request channel and non-blocking
response channel); a separate virtual channel is used for
adaptive routing with the ability to switch to a nonblocking
deterministic channel in case of potential deadlock;
two more virtual channels are used to send messages
over a virtual subnet for collective operations. The collective
operations (broadcast and reduce) support is implemented
using a virtual subnet with a tree topology, which
is applied to multidimensional torus topology.
Deterministic routing preserves the order of packet transmission and prevents deadlocks; adaptive routing uses one of the possible minimum routes for delivering packets, ignoring the order of directions, which allows to bypass the overloaded and broken parts of the network. 

The link layer supports fault tolerant transmission of packets using packet numbering, counting for each checksum and retransmission if the checksum recorded in the last flit of the packet does not coincide with that calculated after the transfer. There is also a mechanism for bypassing failed communication channels and nodes by rebuilding the routing tables and using a non-standard first / last steps of the packet route. For performing various service operations, including setting up / rebuilding routing tables, and performing some calculations, a service processor can be used that interacts with the adapter via the ELB interface.

\begin{figure}[h]
	\begin{center}
		\includegraphics[height=4cm]{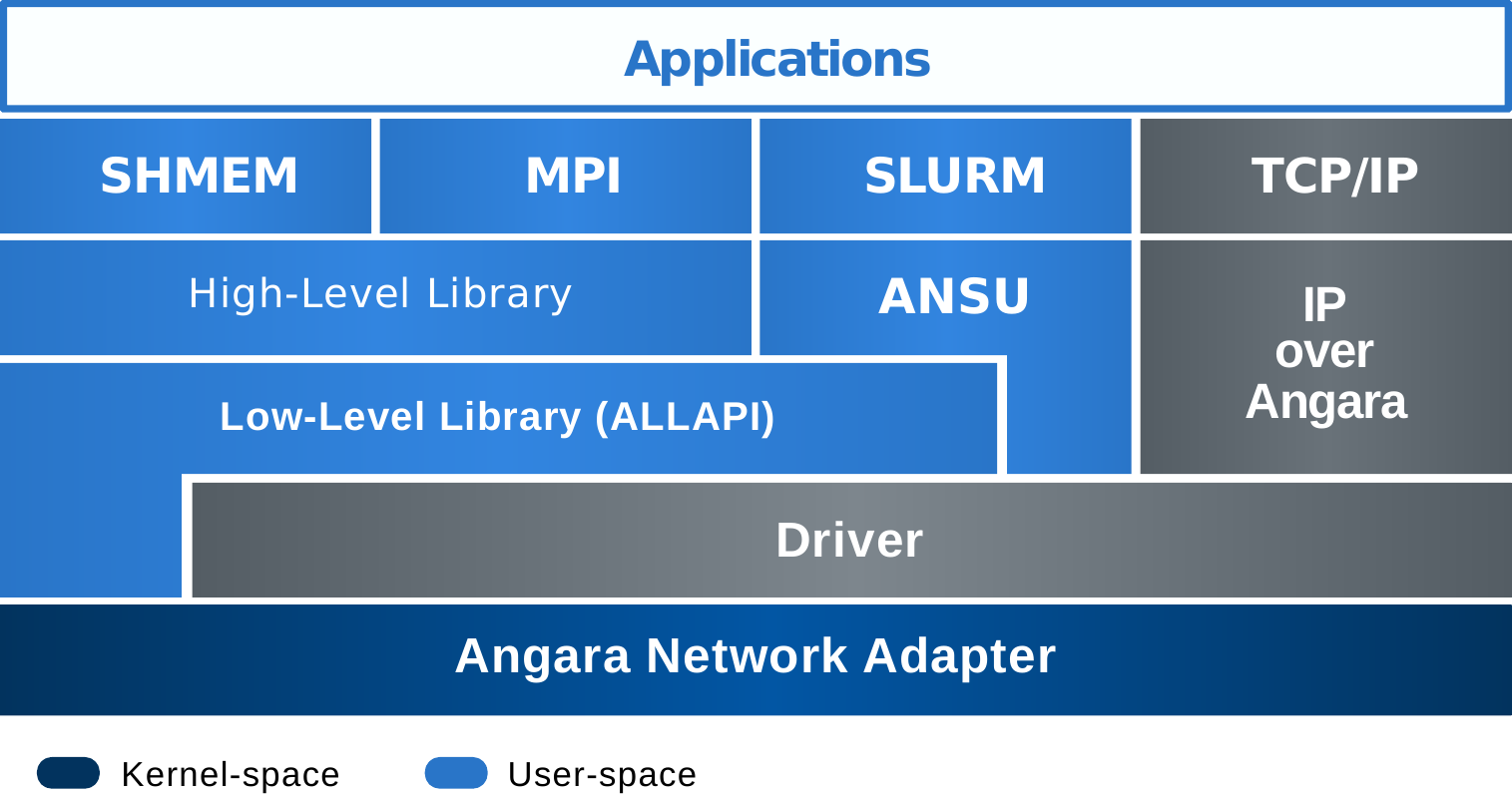}
		\caption{The Angara software stack.}
		\label{fig:software}
	\end{center}
\end{figure}

The Angara chip supports simultaneous operations with
multiple threads/processes of a user task; it is implemented as
several injection channels available for use by independent
packet buffers. 

The network adapter at the hardware level supports remote (RDMA) write, read, and atomic operations. Atomic operations of two types are available -- addition and exclusive OR. Each node
has a dedicated memory region available for remote access
from other nodes to support OpenSHMEM and PGAS. Different programming models are supported, including
MPI and OpenMP. The Angara software stack is presented in Figure \ref{fig:software}.

The network adapter extension card can be connected to
the card in adjacent nodes by up to six cables (or up to eight
with an extension card). The following topologies are supported:
a ring, 2D, 3D and 4D torus.

\subsection{Deterministic Routing}

Deadlock-free deterministic routing in the Angara interconnect is performed preserving the order of
directions (Direction order routing, DOR) \cite{dally1986torus}. All directions in a network route are used in a predetermined order $+X+Y+Z+K-X-Y-Z-K$.
Direction order algorithms have a property of absence of mutual locks between the rings of torus topology with any number of simultaneous requests for data transmission over the network. To avoid deadlocks in a ring (movement without changing a direction) the bubble routing rule is used \cite{adiga2005blue,scott1996cray,puente1999adaptive}.

The second Angara routing restriction is a \textit{direction bit rule}, which does not allow both positive and negative directions of a dimension in a route. $D_{dirbit}$ denotes a set of directions matches the direction bit rule.

The First Step / Last Step (FS/LS) method is used in the Angara interconnect as a mechanism for relaxing routing rules. The FS/LS method provides a possibility of a first positive and/or a last negative non-standard steps in a route. 
A route from $u$ to $v$ nodes with FS/LS steps can be written as $u, D_{FS}, D_{dirbit}, D_{LS}$, where $D_{FS}$ is a positive step, $D_{LS}$ is a negative step. 
Note that a set of directions $D_{FS}, D_{dirbit}, D_{LS}$ can violate the direction bit and the direction order rules, a deadlock can be arised. 

Angara system software includes a SLURM plugin called Angara Node Selection Utility, ANSU. Each router in a network node contains a routing table, which is written by ANSU before each user application run. 
The described deterministic Angara routing rules allow path diversity, i.e. multiple routes between a source-destination pair. Each routing table for each node contains a routing information and determines a route to the node. When a packet is injected to a network, a router creates a 128-bit header flit with a route information from a routing table for a destination node. The header flit is added to a network packet, then the packet deterministically moves through the network.

\section{Routing Graph}

Let $N$ be a set of network nodes. The $n$-dimensional torus has $d_1 \times d_2 \times \cdots \times d_n$ nodes, with $d_j$ nodes along each dimension $j$, where $d_j \geq 2$ for $1 \leq j \leq n$. Each node $u \in N$ is identified by $n$ coordinates $(u_1, u_2, ..., u_n)$, where $0 \leq u_j < d_j$ for $1 \leq j \leq n$. 

The direction $D_j$ is a vector $(0,...,\underbrace{\pm1}_{j\;mod\;n},...,0)$, where each coordinate is zero except a position $j\;mod\;n$ and it is equal $+1$, if $1 \leq j \leq n$ or $-1$, if $n + 1 \leq j \leq 2n$. The direction $D_j$ is \textit{positive}, when $1 \leq j \leq n$ and is \textit{negative}, when $n + 1 \leq j \leq 2n$.

Two nodes $u=(u_1, u_2, ..., u_n)$ and $v$ are neighbors in a direction $D_j$ if $v = u + D_j = (u_1, ..., (u_{j\;mod\;n} \pm 1)~mod~ d_{j\;mod\;n}, ..., u_n), 1 \leq j \leq 2n$. 
A set of all directions is denoted by $\mathcal{D}=\{D_j\}, 1 \leq j \leq 2n$. On the set of directions $\mathcal {D}$ we introduce the order: $D_i < D_j$, if $i < j$.

The channel dependency graph is not suitable for building routes in the Angara interconnect, because it does not support the Angara direction bit routing rule: the channel dependency graph does not allow to analyze an interconnect packet route history, which is needed to control the absense in the packet route of simultaneous using of positive and negative directions of a dimension. 

For Angara routing table generation we propose a special \emph{routing graph} (RG).
The idea of the routing graph is to store a history of a possible packet route with a contracted representation of the history. 

\subsection{First and Last Steps Preserving Direction Order Routing}
\label{sec:rtgraph}

Initially, we consider a case of FS/LS steps that preserve the direction order.
In a routing graph $RG$ for each network node $u^i \in N$ a vertex set $\{U^i_{begin}$, $U^i_{FS_j}$, $U^i_{dirbit_k}$, $U^i_{LS_j}$, $U^i_{end}\}$ is defined according to the following.
\begin{enumerate}
	\item $U^i_{begin}$ is a vertex from which a path is started (injection to the network).
	\item $U^i_{FS_j}$ is a vertex that can be reached by completing the first step $FS_j \in \mathcal{D}$ and $FS_j$ is a positive direction, $1 \leq j \leq n$.
	\item $U^i_{dirbit_k}$ is a vertex that can be reached by completing steps in the $dirbit_k$ ordered direction list, $1 \leq k \leq 3^n - 1$. The ordered direction list $dirbit_k$ satisfies the direction bit rule, then $dirbit_k$ can be presented as a $n$-dimension vector, in which for each network dimension a packet might do 3 kind of steps: in a positive direction, in a negative direction or no steps. 
	\item $U^i_{LS_j}$ is a vertex that can be reached by completing the last step $LS_j \in \mathcal{D}$ and $FS_j$ is a negative direction, $n + 1 \leq j \leq 2n$.
	\item $U^i_{end}$ is a vertex in which a path is finished (ejection from the network).
\end{enumerate}

\begin{figure}[h]
	\begin{center}
		\includegraphics[height=8cm]{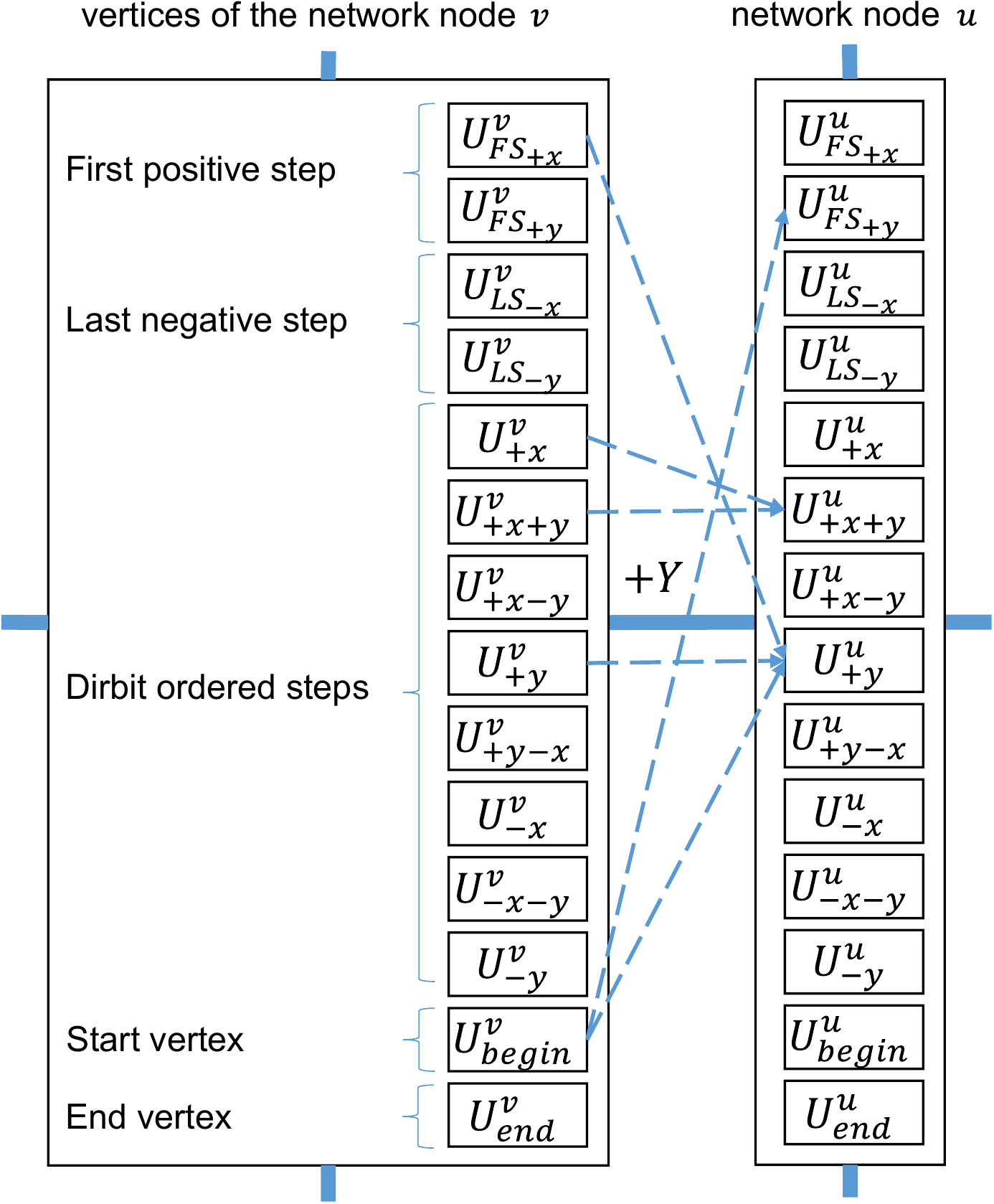}
		\caption{Subgraph example of a routing graph for a 2D torus topology system.}
		\label{fig:routinggraph}
	\end{center}
\end{figure}

Vertices of a $RG$ are connected in such a way that a movement from a vertex to another vertex corresponds to packet route through the corresponding network nodes and satisfies the Angara routing rules. The directed edge set of a routing graph $RG$ is defined according to the following possible steps.

\begin{enumerate}
	\item Consider a vertex $U^i_{begin}$ corresponding to a network node $u^i$, which is a start vertex for a path. A first step of a route from $u^i$ can be
		\begin{itemize}
			\item a positive non-standard step in a $D_k$ direction to a vertex $U^j_{FS_k}$, which corresponds to a node $u^j = u^i + D_k$
			\item a step in a $D_k$ direction to a vertex $U^j_{dirbit_l}$, where $dirbit_l$ corresponds to one single direction $D_k$ and a node $u^j = u^i + D_k$ 
			\item a step to a $U^i_{end}$ vertex, which denotes the end of a path.
		\end{itemize}
	\item Consider the $U^i_{FS_l}$ vertices, $1 \leq l \leq n$, which correspond to a node $u^i$. From the $U^i_{FS_l}$ vertices it can be possible to make a step
	\begin{itemize}
		\item to the $U^j_{dirbit_l}$ vertices, which correspond to a step $u^j = u^i + D_k$ in a $D_k$ direction, in case of $D_l < D_k$. The $dirbit_l$ index corresponds to one single $D_k$ direction.
		\item to a $U^i_{end}$ vertex, the end of a path.
	\end{itemize}
	\item Consider the $U^i_{dirbit_l}$ vertices corresponding to the routes through a node $u^i$ with the ordered direction list $dirbit_l$. From the $U^i_{dirbit_l}$ vertices it can be possible to make a step
	\begin{itemize}
		\item to a $U^j_{dirbit_t}$ vertex in a direction $D_k$, in case of a last direction of $dirbit_l \leq D_k$ and $dirbit_t = \{dirbit_l, D_k\}$, $u^j = u^i + D_k$
		\item to a $U^j_{LS_k}$ vertex in a $D_k$ direction, in case of a last direction of $dirbit_l < D_k$ and $u^j = u^i + D_k$
		\item to a $U^i_{end}$ vertex, the end of a path.
	\end{itemize}
	\item Consider the $U^i_{LS_k}$ vertices corresponding to the routes through a node $u^i$ with a last negative non-standard step in a direction $D_k$. From the $U^i_{LS_k}$ vertices it can be possible to make a step to a $U^i_{end}$ vertex, the end of a path.
\end{enumerate}

Figure \ref{fig:routinggraph} presents a subgraph example of a routing graph for a 2D torus topology system. The subgraph includes two set of vertices for nodes $v$ and $v$ and edges corresponding to a step from $v$ to $u$ along direction $+Y$ .

For each network node $u^i \in N$ a number of vertices $C_V(n) = |U_*^i| = 3^n + 2n +1$. Consider an edge set for a vertex set $U_*^i$. First, the $U_{begin}^i$ vertices have $n + 2n + 1$ edges. Secondly, for each non-standard step $D_l$ there are $2n-j$ edges, the total is $\sum\limits_{j=1}^n(2n-j) = \frac{3n^2-n}{2}$. Thirdly, for each $U^i_{dirbit_l}$ of $3^n - 1$ vertices there are at most $2n$ edges. Finally, we have $n$ edges. A total edge number for each network node $u^i \in N$ is $C_E(n) = |E^i_*| = 2n3^n + 1.5n^2 + 1.5n + 1$.

\begin{theorem}\label{th:1}
	From a network node $u^i \in N$ to a node $u^j \in N$ there is a route if and only if there is a path from a vertex $U^i_{begin}$ to a vertex $U^j_{end}$ in a routing graph $RG(V, E)$, which corresponds to the network topology.
\end{theorem}

The proof is based on a routing graph constuction algorithm. For each step $(u^k, u^l)$, $u^k, u^l \in N$ of a route in a network there is an edge between vertices $U_{ss1}^k$ and $U_{ss2}^l$ in a corresponding routing graph, where $ss2 = ss1$ if the step does not change a route direction or $ss1$ switches to $ss2$ according to the step, if the step changes a route direction. Conversely, for a path in the routing graph there is a route in the network.

\subsection{Deadlock-Free Routing Graph}

In \cite{mukosey2019extended} we relax the direction order rule by augmenting the edge set of the routing graph by adding the edges, which do not lead to deadlocks. We determine these edges by a channel dependency graph. The node set of a channel dependency graph $CDG(C,E)$ consists of the communications channels that can be presented for a torus topology as a pair $(u, D_i)$, where $u \in N$ is a network node, $D_i \in \mathcal{D}$ is a direction.
The edge set $E$ of $CDG$ can be presented as channel pairs $[\cdotp, \cdotp]$, i.e. $[(v, D_i), (u, D_j)] \in E$, where $(v, D_i), (u, D_j) \in C$, $D_i, D_j \in \mathcal{D}$, and $u = v + D_i$.

For a given torus topology we construct a routing graph $RG$ and a channel dependency graph $CDG$, which provides the direction order rule satisfaction. 

To evaluate a possibility of making steps with violation of the deadlock-freedom rules, we build for each vertex of a channel dependency graph a used direction set in the paths from this vertex. The \textit{used direction set} $B(u,D_u)$ is a set of directions ${D_1, ..., D_k}, D_i \in \mathcal{D}, \forall i \; 1 \leq i \leq k$, which can be used in all possible paths started from a vertex $(u,D_u) \in C$ in the channel dependency graph.

\begin{theorem}\label{prop:1}
	New edge $[(u_j, D_j), (u_k, D_k)] \in E$ of $CDG(C, E)$, where $D_j > D_k$ and $u_k = u_j + D_j$ does not create cycles in the graph if $D_j \notin B(u_k, D_k)$.
\end{theorem}

By contradiction, we add the edge $[(u_j, D_j), (u_k, D_k)]$ to the $CDG$ and suppose there is a path in the $CDG$ that creates a cycle. Then there is a path from $(u_k, D_k)$ to $(u_j, D_j)$, then direction $D_j \in B(u_k, D_k)$, it contradicts to the theorem condition. 

Algorithm for adding edges to the channel dependency graph $CDG$ can be divided into two stages. In the first stage the breadth-first search (BFS) algorithm is used to build a used direction set for each $CDG$ vertex. BFS starts from each vertex and stores all directions from the reachable vertices.

In the second stage the algorithm finds possible edges satisfying the Theorem~\ref{prop:1}. For each $CDG$ vertex $(u, D_i)$ the algorithm adds edges to all the neighbor vertices $(v, D_j)$, when $D_i$ is not in the used direction set of $(v, D_j)$. Then we rebuild the used direction set for each $CDG$ vertex from which there is a path to the vertex $(u, D_i)$ by starting BFS from $(v, D_j)$ in $CDG$ with the reverted edges and adding directions from $B(v, D_j)$ to all reachable vertices.
The second stage are performed until all possible edges are constructed.

After channel dependency graph analysis we need to augment the routing graph. Each added edge $[(u_i, D_i), (u_j, D_j)]$ of $CDG$ allows a new route turn that violates the direction order rule, i.e. $D_i > D_j$. The Angara routing allows the direction order rule violation only for a first positive and/or a last negative non-standard steps. When $D_i$ is a first positive step, for a corresponding routing graph vertex $U^j_{FS_i}$ we add an edge to a vertex $U^k_{D_j}$ with $u_k = u_j + D_j$. When $D_i$ is a last negative step, for the corresponding $U^j_{dirbit}$ vertices, such that $D_i$ is a last direction in the ordered direction list ${dirbit}$, we add an edge to a vertex $U^k_{LS_j}$ with $u_k = u_j + D_j$.

The added routing graph edges preserve Theorem \ref{th:1} correctness.

\section{Routing Algorithms}	

In this section we present the previously proposed BFS based algorithm, devised Genetic and SSSP algorithms.

\subsection{BFS Algorithm}

A BFS routing table generation algorithm is based on the breadth-first search algorithm in a routing graph. 
The paths are built by iterative application of the breadth-first search algorithm, see Algorithm \ref{BFSroutes}. The edge link attribute denotes a link corresponding to an edge, we note that for an edge set there is a link. The reverse path is used to generate the routing tables, called $routes$. All edge weights along the built paths are increased by 1. Ascending vertex order in $Q$ by outgoing edge weights provides the better load balancing. \FIXME{Complexity?}

The whole routing BFS algorithm, see Algorithm \ref{allBFSroutes}, iterates over all nodes in the given set $N_a$ to find the paths from a source
to all other nodes. The next source node is the most distant from the current node (a GET\_MAX\_REMOTE procedure), this heuristic provides the better load balancing.
Initially all edge weights are set to 0. The breadth-first search in the routing graph $RG = G(V,E)$ has a complexity of $O(|V|+|E|)$, the time complexity of the BFS routing algorithm is $O(|N| \cdot (|V|+|E|)) = O(|N|^2 \cdot [C_V(n)+C_E(n)])$, where $C_V(n)$ and $C_E(n)$ are relatively large constants, we note that these constants depend on number of torus dimensions $n$.

	
\begin{algorithm}
	\caption{Build routes by BFS in a routing graph}\label{allBFSroutes}
	\hspace*{\algorithmicindent} \textbf{Input:} $RG$ -- routing graph, $N$ -- node set\\
	\hspace*{\algorithmicindent} \textbf{Output:} Routing table for $N$\\
	\begin{algorithmic}[1]
		\Procedure{build\_rt\_bfs}{$RG, N$}
		\State $v_{source} = RG$.beginVertex($N$.first())
		\Repeat
			\State \textsc{build\_bfs\_routes}($RG$, $v_{source}$)
			\State mark\_processed($v_{source}$)
			\State $v_{source}$ = \textsc{get\_max\_remote}($N, v_{source}$)
		\Until{not all\_processed($N$)}
		\EndProcedure \\
		
		\Procedure{get\_max\_remote}{$N$, $v_{last}$}
		\State $v_{max\_remote}$ = $v_{last}$
		\For{\textbf{all} $v$ $\in$ not\_processed($N$)}
			\If{length($v_{last}$, $v_{max\_remote}$) $<$ \\           length($v_{last}$, $v$)}
				\State $v_{max\_remote}$ = $v$
			\EndIf
		\EndFor
		\State \Return $v_{max\_remote}$
		\EndProcedure
		
	\end{algorithmic}	
\end{algorithm}

\begin{algorithm}
	\caption{Build routes by BFS in a routing graph from a source vertex}\label{BFSroutes}
	\hspace*{\algorithmicindent} \textbf{Input:} Routing graph $RG(V,E)$, start vertex $source$ \\
	\hspace*{\algorithmicindent} \textbf{Output:} All shortest paths from $source$
	
	\begin{algorithmic}[1]
	\Procedure{build\_bfs\_routes}{$RG$, $source$}
			\State /* \textit{build BFS tree from source to all targets} */
		\For{\textbf{all} $v \in V$} 
				initialize $v$ ($v$.parent = $\emptyset$)
		\EndFor 
		\State $Q$ = $\{source\}$
		\While{$Q \neq \{\}$}
			\State $Q_{next}$ = $\{\}$
			\State /* \textit{ascending vertex sort in Q by outgoing edge weights} */
			\For{\textbf{all} vertex $u \in$ ascending sort $(Q)$}
				\For{\textbf{all} $(u, v) \in E$}
					\If{$v$.parent $== \emptyset$}
						\State $v$.parent = $u$
						\State $Q_{next}$ = $Q_{next}$ $\cup$ $\{v\}$
					\EndIf			
				\EndFor
			\EndFor
			\State $Q = Q_{next}$
		\EndWhile
		\State /* build paths from BFS tree in reverse order */
		\For{\textbf{all} end vertex $dst$ $\in$ $RG.V^*_{end}$}
			\State $u = dst$
			\Repeat
			\State $routes$[$source$.node, $dst$.node]. \\ 
			\hspace*{6.2em} push\_back($u$.node)
			\For{\textbf{all} $(u_1$, $u_2$) = $e \in E$, \\
			\hspace*{6.2em} where $e$.link == ($u$.parent, $u$).link}
			\State $e$.weight = e.weight $+ 1$
			\EndFor
			\State $u = u$.parent
			\Until{$u \neq source$}
		\EndFor
	\EndProcedure	
	\end{algorithmic}
\end{algorithm}

\subsection{Genetic Algorithm}

We design a Genetic algorithm for heuristic search space exploration. We consider the Genetic algorithm as an alternative way to evaluate an affordable routing table quality. The Genetic algorithm operates with a population of solution candidates represented as chromosomes. Then, based on current generation, it produces the next generation by applying crossover on some randomly selected chromosome pairs or mutates some random ones and then it selects the best chromosomes based on a fitness function to produce the next generation. Details of this algorithm are as follows.

Chromosome is a set of genes, representing a routing table for each pair of network nodes.
A gene $G^{l}_{k}$ is a route between a pair of network nodes, where $k$ is a position of the gene in a chromosome, because network node pairs can be mapped to positive integers, $l$ -- variant route number for the network node pair. We construct an initial generation by randomly
producing each of 100 chromosomes. We use a two-point crossover operation with panmixia as a parent selection algorithm. In a mutation operation each children gene is randomly changed with probability 2\%.
As a fitness function we use $\sigma(k,\mathcal{R})$ with $k=4$, we identify that it provides rapid convergence. The Genetic algorithm stops when it can not improve the best chromosomes after a rather large number of generations. In our simulation Genetic is terminated after 30 consecutive iteration or if the fitness value of next best solution less then 0.05 of fitness value of global best solution.

\subsection{SSSP Algorithm}

The third routing table generation algorithm is presented in Algorithm \ref{alg:allSSSProutes} and is based on a single-source shortestpath (SSSP) algorithm in a routing graph. 

In our algorithm we aim to reduce a number of SSSP calls, preserving the desired balancedness quality.
First, we count a path number and a path length from each network node to all other network nodes by the breadth-first search algorithm in the routing graph $RG$. We detect that for the 49 2D, 339 3D, 1173 4D random torus topology systems, where each dimension size $d_i \in [2; 8]$, 81.6\%, 58.2\% and 36.6\% source-destination node pairs, on average, respectively, have a single minimum route, we assume the Angara routing rules. Then we apply these unique routes and adjust edge weights.

Secondly, we sort and group source-destination node pairs by ascending order of a route turn number in torus topology and then by ascending order of route length. Also we group node pairs by a source node in order that we can run a BUILD\_SSSP Algorithm \ref{alg:SSSP} from a source to a set of destination nodes. Then for the sake of the best balancing quality we apply those built paths that do not cross each other, adjust edge weights and repeat the BUILD\_SSSP call for a set $N_{dst}$ of remaining destination nodes until $N_{dst}$ is not an empty set.  

Algorithm \ref{alg:SSSP} builds single-source shortest paths from a source to a given set $N_{dst}$. An initial edge weight of routing graph is $|N|^2$ to force minimal shortest paths following T. Hoefler in \cite{domke2011deadlock}. Only $|N|\cdot|N-1|$ paths are considered, each $w_e^i<|N|^2$, from this it follows that the shortest-path algorithm with the edge initialization never chooses a detour. 

\begin{figure}[h]
	\begin{center}
		\includegraphics[height=6cm]{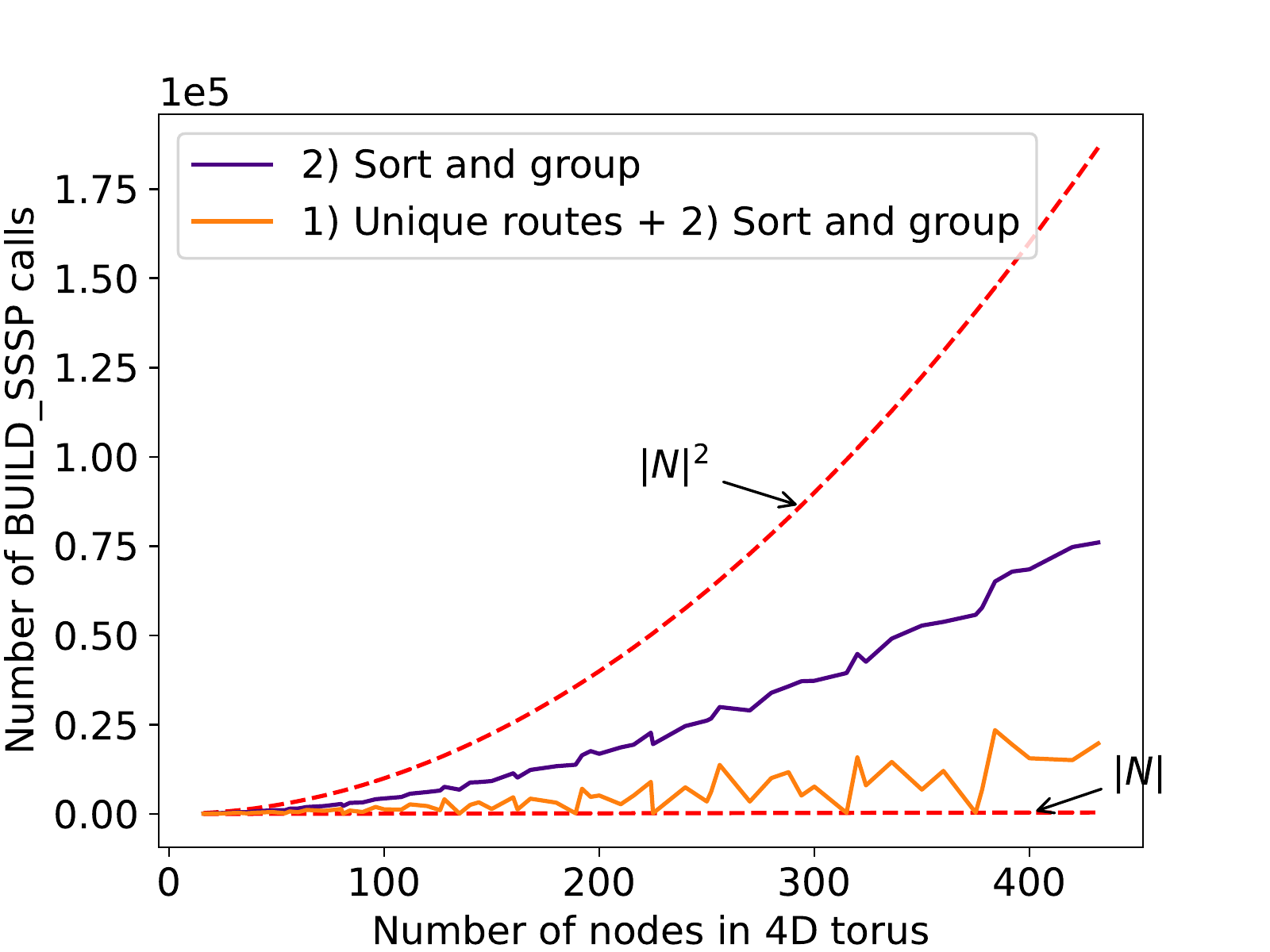}
		\caption{Number of BUILD\_SSSP calls within the SSSP routing algorithm.}
		\label{fig:sssp_cals_number}
	\end{center}
\end{figure}

We identify that the edge initialization allows to consider vertices in each breadth-first search level in arbitrary order. When a level processing is finished, then tentative distance of a vertex belonged to this level is settled. Therefore we use actually the breadth-first search algorithm with edge relaxation instead of the Dijkstra algorithm and thus optimize a complexity of the proposed algorithm to $O(|N|\cdot [C_V(n)+C_E(n)])$ instead of $O(|N|\cdot log|N|\cdot [C_V(n)+C_E(n)])$. Also the proposed algorithm breaks when all $N_{dst}$ vertices are processed.

The complexity of the SSSP routing algorithm is dominated by the second algorithm stage. Figure \ref{fig:sssp_cals_number} presents a call number of the BUILD\_SSSP algorithm. If a routing algorithm consists of only the second stage called 'Sort and group', then a number of BUILD\_SSSP calls is between $|N|$ and $|N|^2$. Adding the first stage called 'Unique routes' allows to significantly reduce a number of BUILD\_SSSP calls. The first stage has a complexity $O(|N|^2 \cdot [C_V(n)+C_E(n)])$ as $|N|$ calls of the breadth-first search algorithm. But still an upper-bound complexity of the SSSP routing algorithm is $O(|N|^3 \cdot [C_V(n)+C_E(n)])$.


\begin{algorithm}
	\caption{Build routes by SSSP in a routing graph}\label{alg:allSSSProutes}
	\hspace*{\algorithmicindent} \textbf{Input:} $RG$ -- routing graph, $N$ -- node set\\
	\hspace*{\algorithmicindent} \textbf{Output:} Routing table for $N$\\
	\begin{algorithmic}[1]
		\Procedure{build\_rt\_sssp}{$RG, N$}
		\State /* 1) Unique routes */ 
		\For{\textbf{all} $u \in N$} 
			\State count a path number and a path length from $RG$.beginVertex($u$) to end vertices of all other nodes by BFS
		\EndFor 
		\For{\textbf{all} $(src,dst) \in (N,N) : (src,dst)$. path\_number == 1} 
		\State $routes$[$src$, $dst$] = buildBFSRoute($src$, $dst$)
		\EndFor 
		\State /* 2) Sort and group */ 
		\State $groups$ = sort and group pairs $(src, dst) \in (N,N)$ by 1) ascending order of turn number in a route 2) ascending order of path length 3) $src$
		\For{\textbf{all} $group(src, N_{dst}) \in groups$} 
		\While{$N_{dst} \neq \{\}$}
		\State $paths$ = \textsc{build\_sssp}($RG, src, N_{dst}$)
		\State $N_{b\_dst} = N_{dst} :$ routes corresponding to $paths$ do not cross with each other	
		\State $routes[src, N_{b\_dst}] = $applyRoutes(
		\Statex \hspace{\algorithmicindent} \hspace{\algorithmicindent} \hspace{\algorithmicindent} $paths$[$src, N_{b\_dst}$])
		\State increase edge weight by 1 along all $paths$
		\State $N_{dst}=N_{dst} \setminus N_{b\_dst}$
		\EndWhile
		\EndFor 
		\EndProcedure \\
	\end{algorithmic}	
\end{algorithm}

%

\begin{algorithm}
	\caption{Build single source shortest paths by BFS with relaxation in a routing graph}\label{alg:SSSP}
	\hspace*{\algorithmicindent} \textbf{Input:} Routing graph $RG(V,E)$, start vertex $source$, destination node set $N_{dst}$\\
	\hspace*{\algorithmicindent} \textbf{Output:} All shortest paths from $source$ to $N_{dst}$
	
	\begin{algorithmic}[1]
		\Procedure{build\_sssp}{$RG$, $source$, $N_{dst}$}
		\State /* \textit{build a SSSP tree from source to all targets} */
		\For{\textbf{all} $v \in V$} 
		\State initialize $v$ ($v$.distance = $\infty$, $v$.parent = $\emptyset$)
		\EndFor 
		\State $Q$ = $\{source\}$, $v_{processed} = 0$
		\While{$Q \neq \{\}$}
		\State $Q_{next}$ = $\{\}$
		\For{\textbf{all} vertex $u \in$ $Q$, \textbf{all} $(u, v) \in E$}
		\If{$u$.distance + $w_e < v$.distance}
		\State \textbf{if} $v$.distance == $\infty$ \textbf{then} \Statex \hspace{\algorithmicindent} \hspace{\algorithmicindent} \hspace{\algorithmicindent} \hspace{\algorithmicindent} $v_{processed} =v_{processed}+1$
		\State $v$.distance = $u$.distance  + $w_e$
		\State $v$.parent = $u$
		\State $Q_{next}$ = $Q_{next}$ $\cup$ $\{v\}$
		\EndIf			
		\EndFor
		\State 	\textbf{if} $v_{processed} == |N_{dst}|$ \textbf{then break}
		\State $Q = Q_{next}$
		\EndWhile
		\EndProcedure	
	\end{algorithmic}
\end{algorithm}

\section{Experimental Evaluation} 

We implemented the BFS, Genetic and SSSP algorithms using C language as a part of a SLURM plugin called Angara Node Selection Utulity (ANSU). We compare the different routing table generation algorithms by the balancedness quality and runtime on a synthetic system set and benchmark the performance for the communication patterns and NPB benchmarks on the Desmos cluster.

\subsection{Balancedness and Runtime}

\begin{figure*}
	\begin{center}		
		\subfloat[2D]{
			\label{fig:rt2D}
			\includegraphics[width=0.33\textwidth]{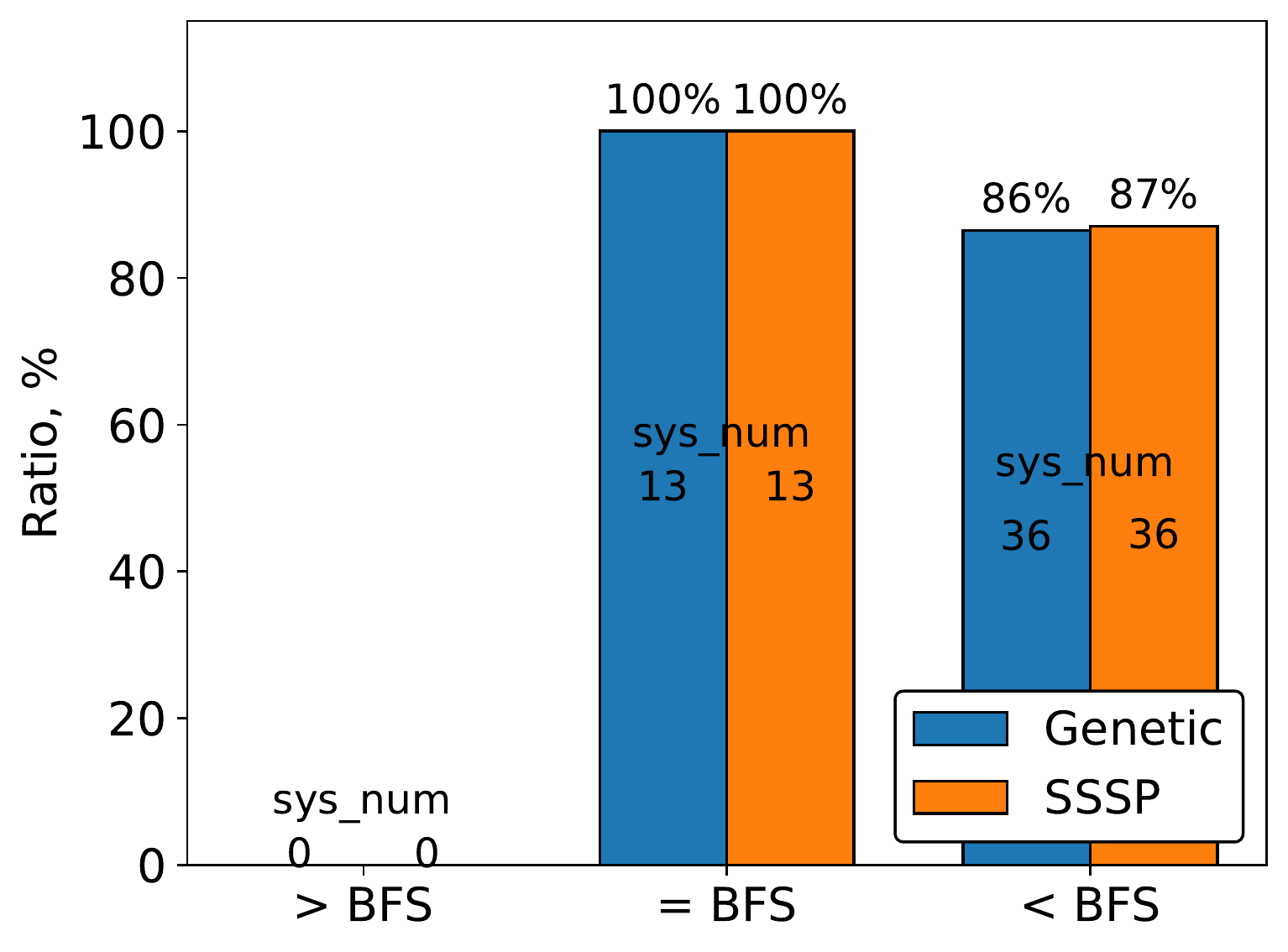}
		}
		\subfloat[3D]{
			\label{fig:rt3D}
			\includegraphics[width=0.33\textwidth]{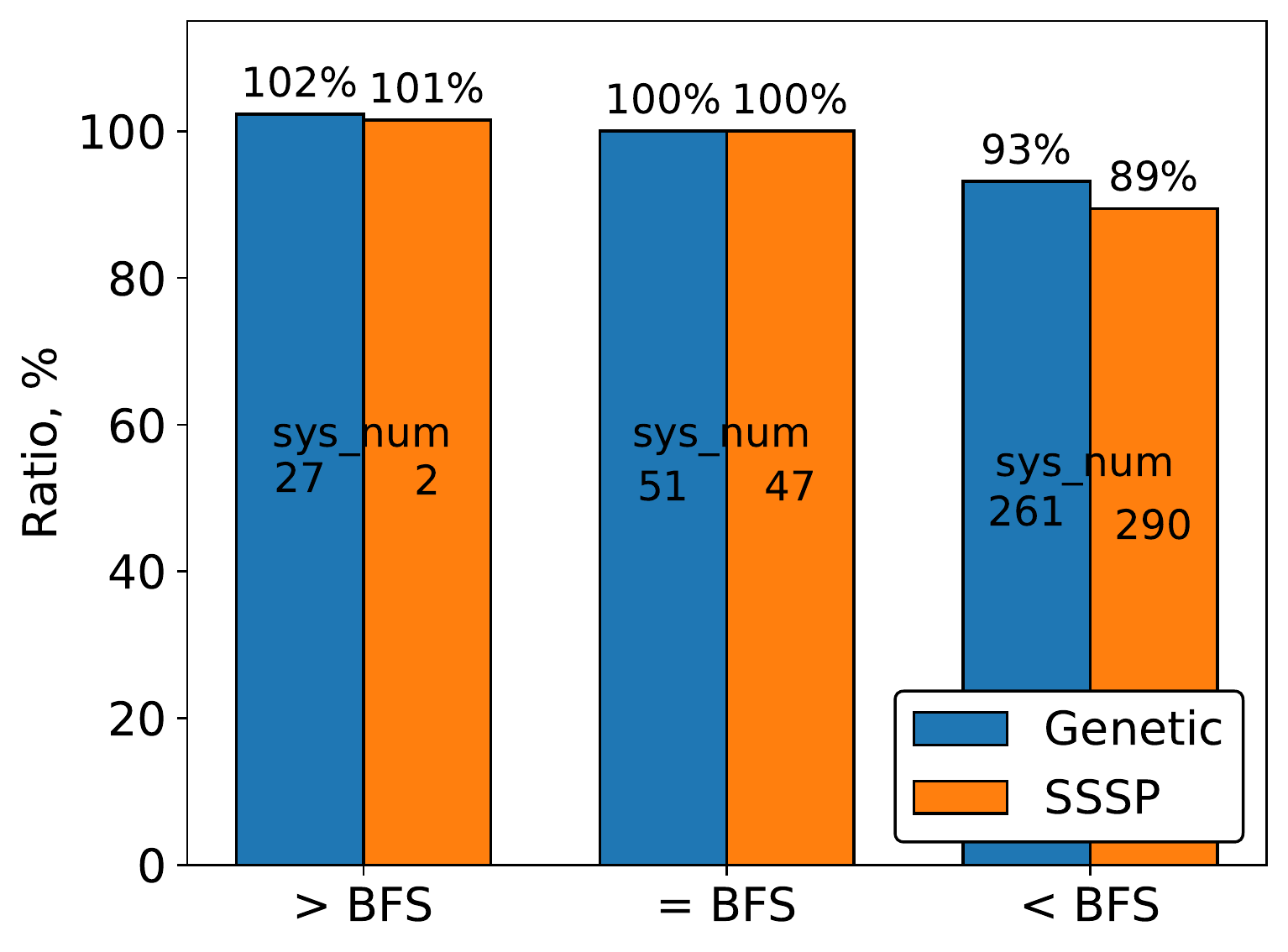}
		}
		\subfloat[4D]{
			\label{fig:rt4D}
			\includegraphics[width=0.33\textwidth]{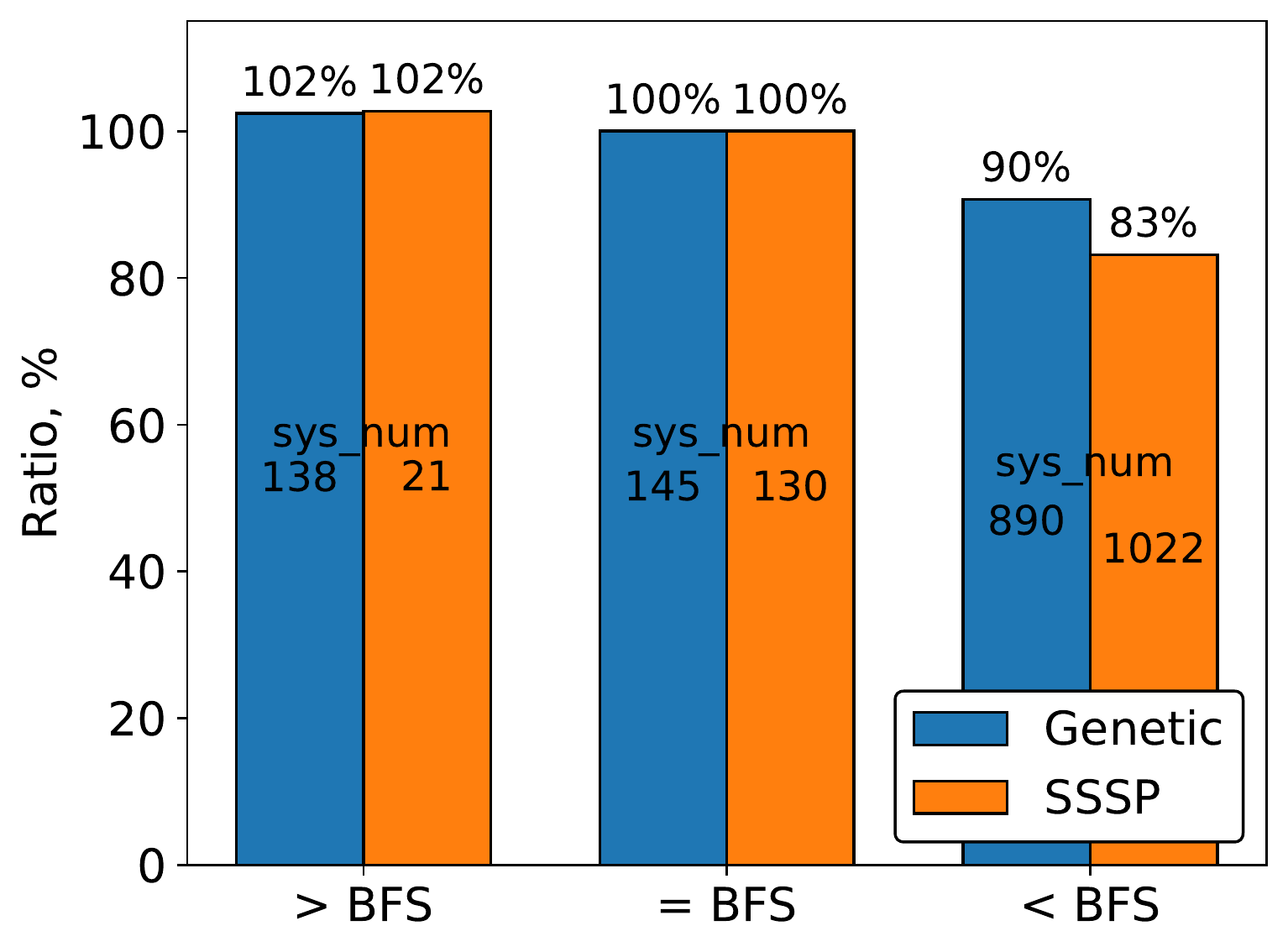}
		}
	\hfil
		\subfloat[2D]{
	\label{fig:runtime2D}
	\includegraphics[width=0.33\textwidth]{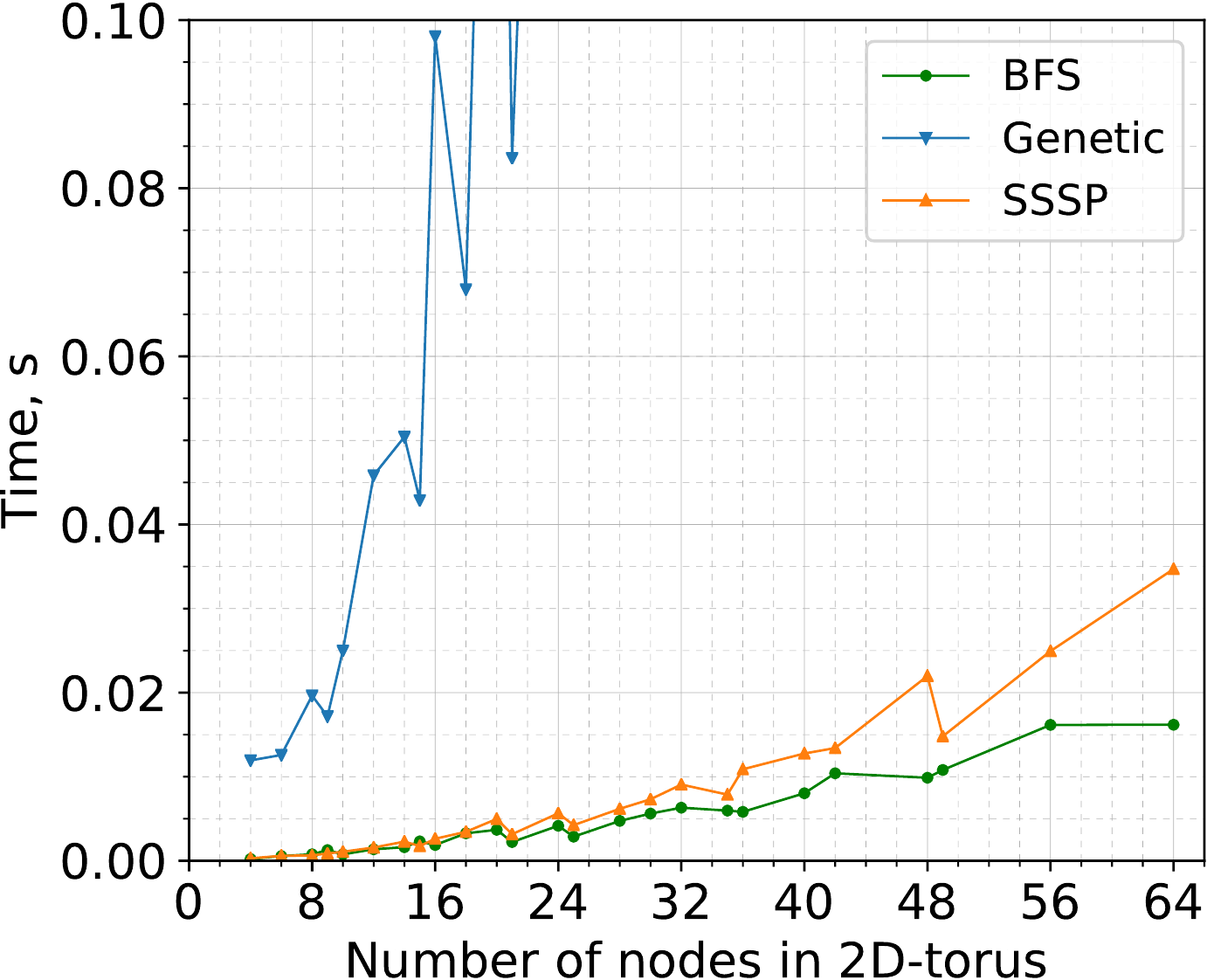}
}
\subfloat[3D]{
	\label{fig:runtime3D}
	\includegraphics[width=0.33\textwidth]{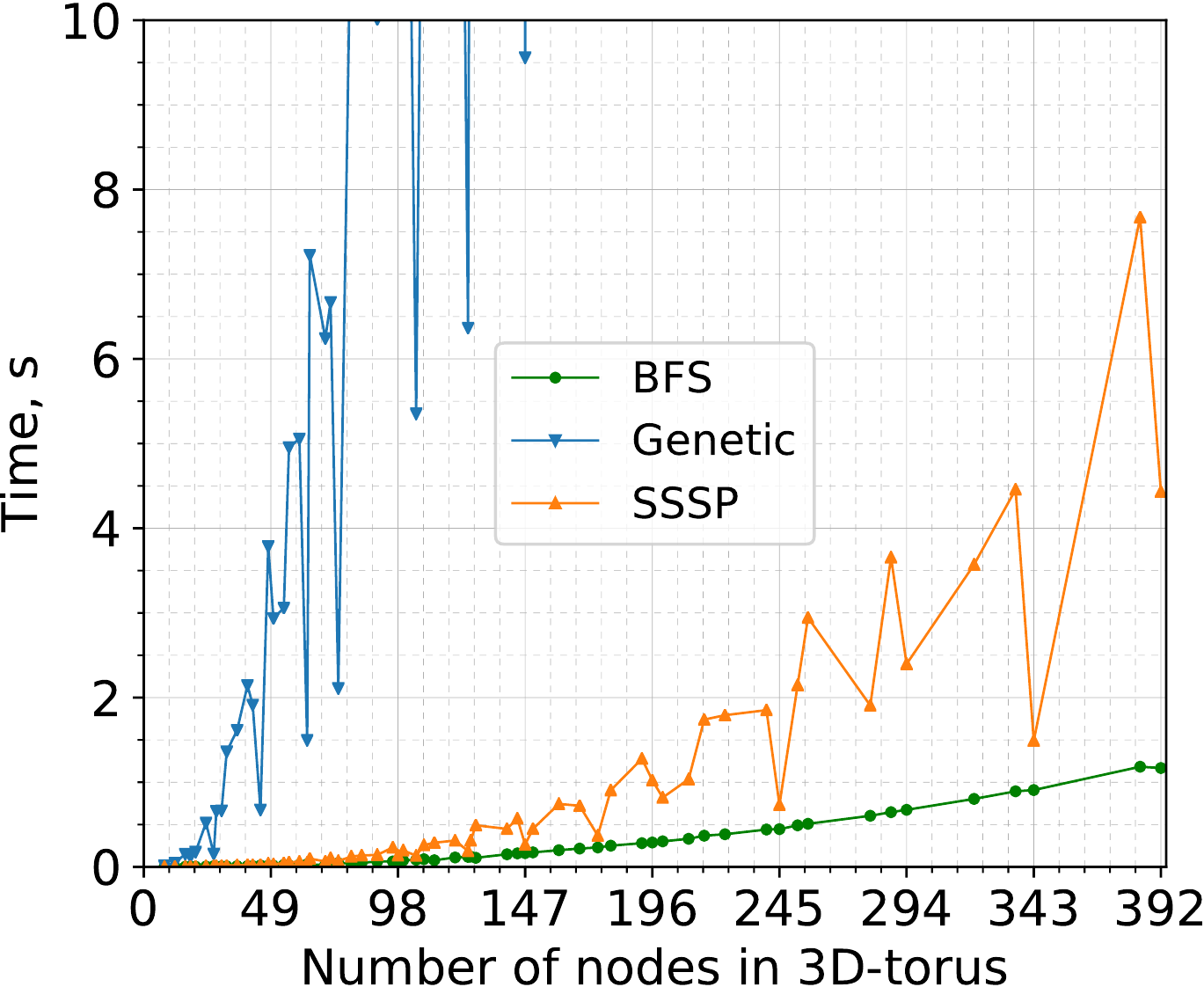}
}
\subfloat[4D]{
	\label{fig:runtime4D}
	\includegraphics[width=0.33\textwidth]{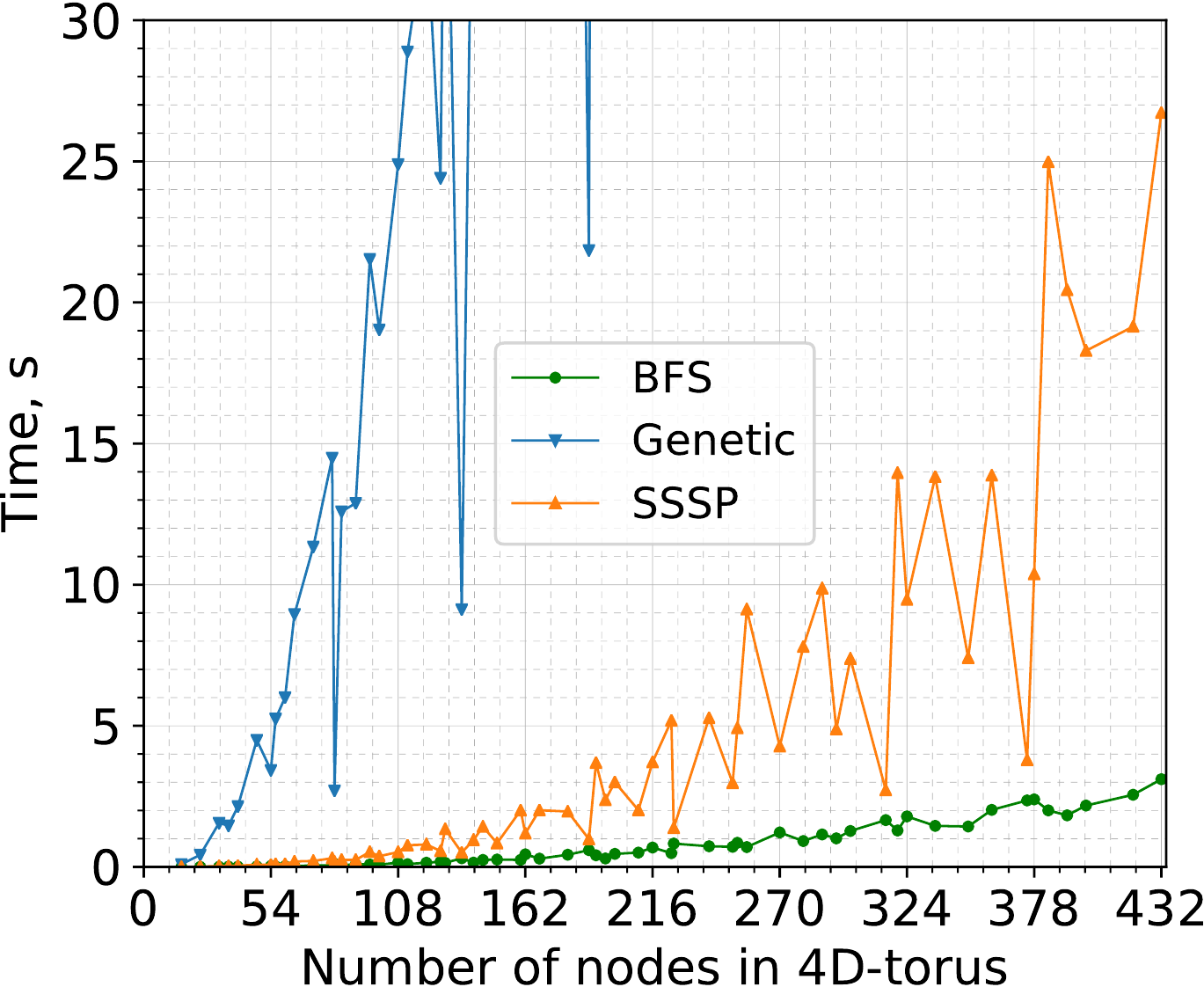}
}	
	\end{center}
	\caption{Comparison the edge-forwarding index ratio with respect to the BFS algorithm and runtime of routing algorithms on 2D, 3D and 4D torus topologies.}
	\label{fig:rtalgs}
\end{figure*}

We consider 49 2D, 339 3D, 1173 4D torus random topology systems, each dimension size $d_i \in [2; 8]$. When $d_i$ is 2, then along this dimension a system topology is mesh, torus in other cases.

Figure \ref{fig:rtalgs} compares the quality and runtime of the presented routing algorithms. The 2D, 3D and 4D torus topology systems we divide into three groups by comparison (less, equal or greater) the edge-forwarding index of the BFS algorithm with the Genetic and SSSP algorithms. For each group we report the ratio of the total edge-forwarding index for the Genetic and SSSP algorithms to the BFS algorithm. Also we report a number of systems for each group (sys\_num in Figure \ref{fig:rtalgs}). The SSSP algorithm outperforms Genetic and BFS in the most cases. In the most cases, on average, SSSP is better than Genetic by 4\% and 7\% for 3D and 4D torus topologies, and is better than BFS by 13\%, 11\% and 17\% for 2D, 3D and 3D torus topologies (see Figures \ref{fig:rt3D} and \ref{fig:rt4D}).

Figures \ref{fig:runtime2D}, \ref{fig:runtime3D} and \ref{fig:runtime4D} presents the algorithm runtime. The evaluation shows that Genetic algorithm is too computationally expensive to be used in practice. Recall (Section \ref{sec:rtgraph}) that the total vertex and edge number in a routing graph substantively depend on the number of dimensions. We define 0.25 sec as a practical boundary threshold of the algorithm runtime, then the SSSP algorithm can be used in practice up to 150 nodes of 3D torus topology, 90 nodes torus topology of 4D and for 2D torus topology systems. The BFS algorithm can be used in practice up to 160 and 300 nodes of 3D and 4D torus topologies, respectively.

\subsection{Real-World System} 

\begin{table}
	\caption{\label{tab:Desmos}The Desmos cluster configuration}
	\begin{center}
		\renewcommand{\arraystretch}{1.2}
		\begin{tabular}{ l c c }
			\hline
			\textbf{Cluster} & \textbf{Desmos} \\
			\hline
			\textbf{Chassis} & SuperServer 1018GR-T \\
			\hline
			\textbf{Processor} & E5-1650v3 (6c, 3.0 GHz) \\
			\hline
			\textbf{GPU} & AMD Radeon Instinct MI50\\
			\hline
			\textbf{Memory} & DDR4 16 GB \\
			\hline
			\textbf{Number of nodes} & 32 \\
			\hline
			\textbf{Interconnect} & Angara 4D torus $4\times2\times2\times2$ \\
			\hline
			\textbf{Operating system} & SLES 11 SP4 \\
			\hline
			\textbf{Compiler} & gcc 7.4.1 \\
			\hline
			\textbf{MPI} & MPICH 3.3  \\
			\hline
		\end{tabular}
	\end{center}
\end{table}
We benchmarked and compared the
routing table generation algorithms on
the 32-node Desmos cluster at Joint Institute for High Temperatures of the Russian Academy of Sciences \cite{stegailov2019angara}. Table \ref{tab:Desmos} provides the Desmos system configuration. All codes are compiled with gcc version 7.4.1 using -O2.

\begin{figure*}
	\begin{center}		
		\subfloat[Transpose]{
			\label{fig:transpose}
			\includegraphics[width=0.4\textwidth]{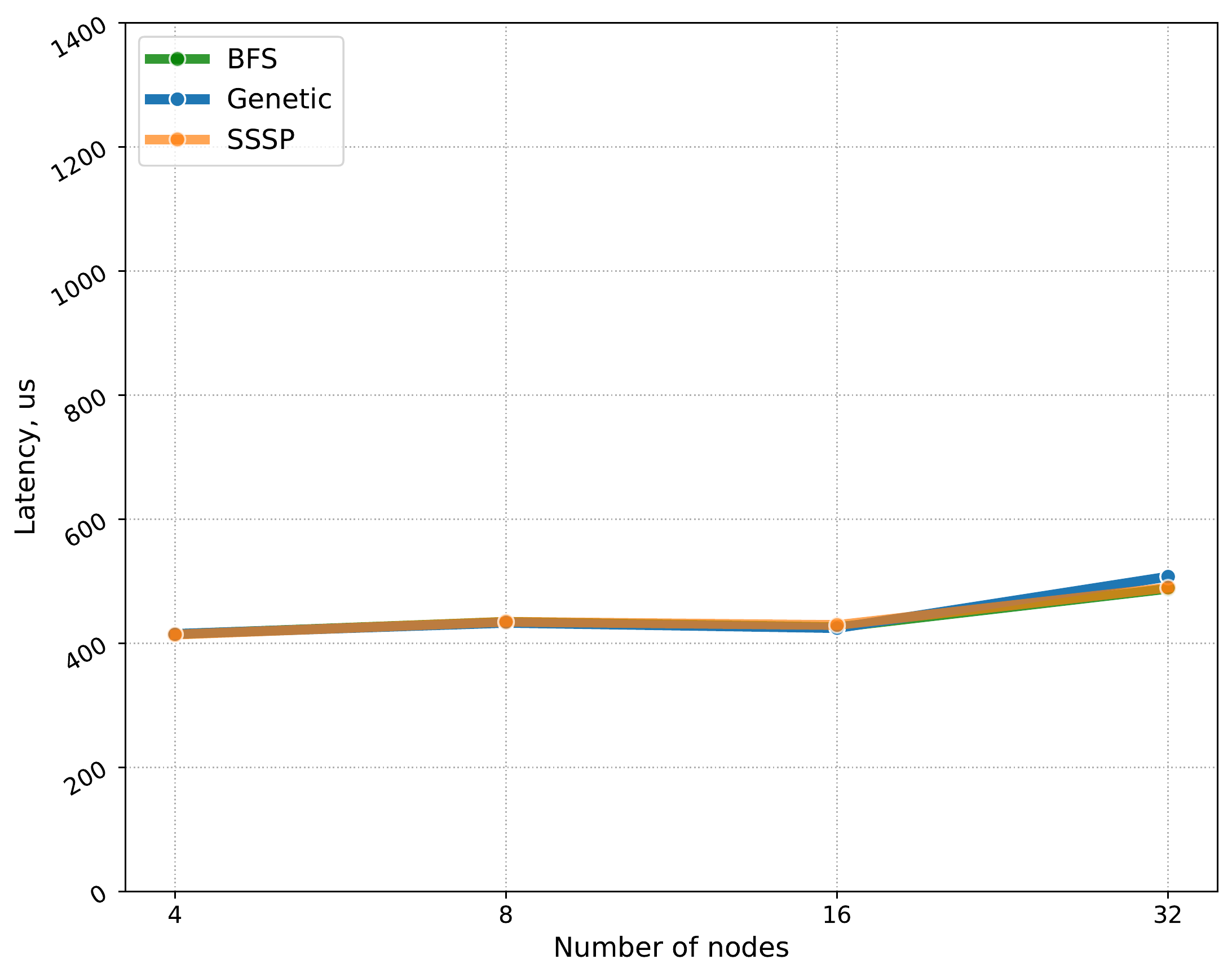}
		}
		\subfloat[Neighbor]{
			\label{fig:neighbor}
			\includegraphics[width=0.4\textwidth]{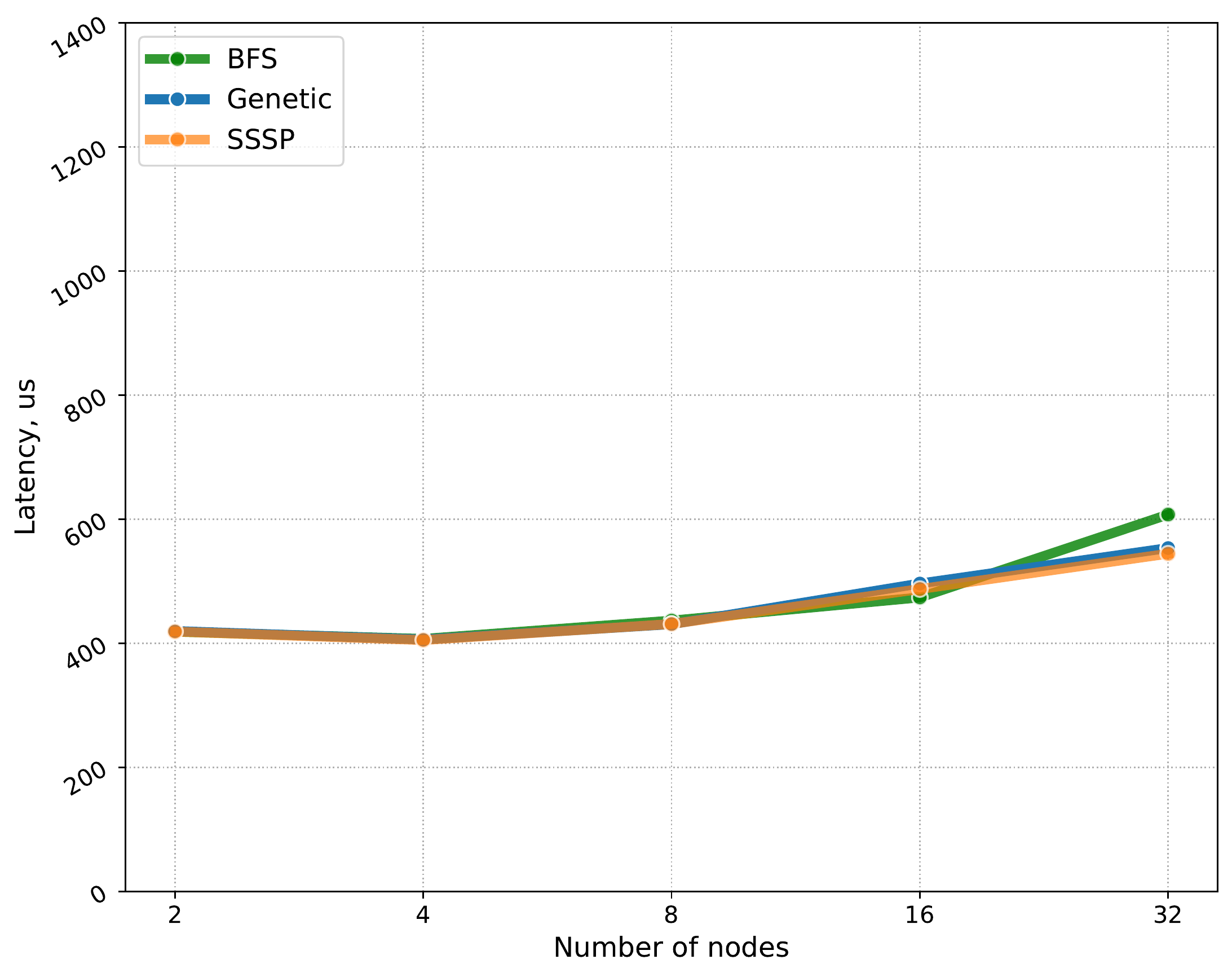}
		}
		\hfil
		\subfloat[Tornado]{
			\label{fig:tornado}
			\includegraphics[width=0.4\textwidth]{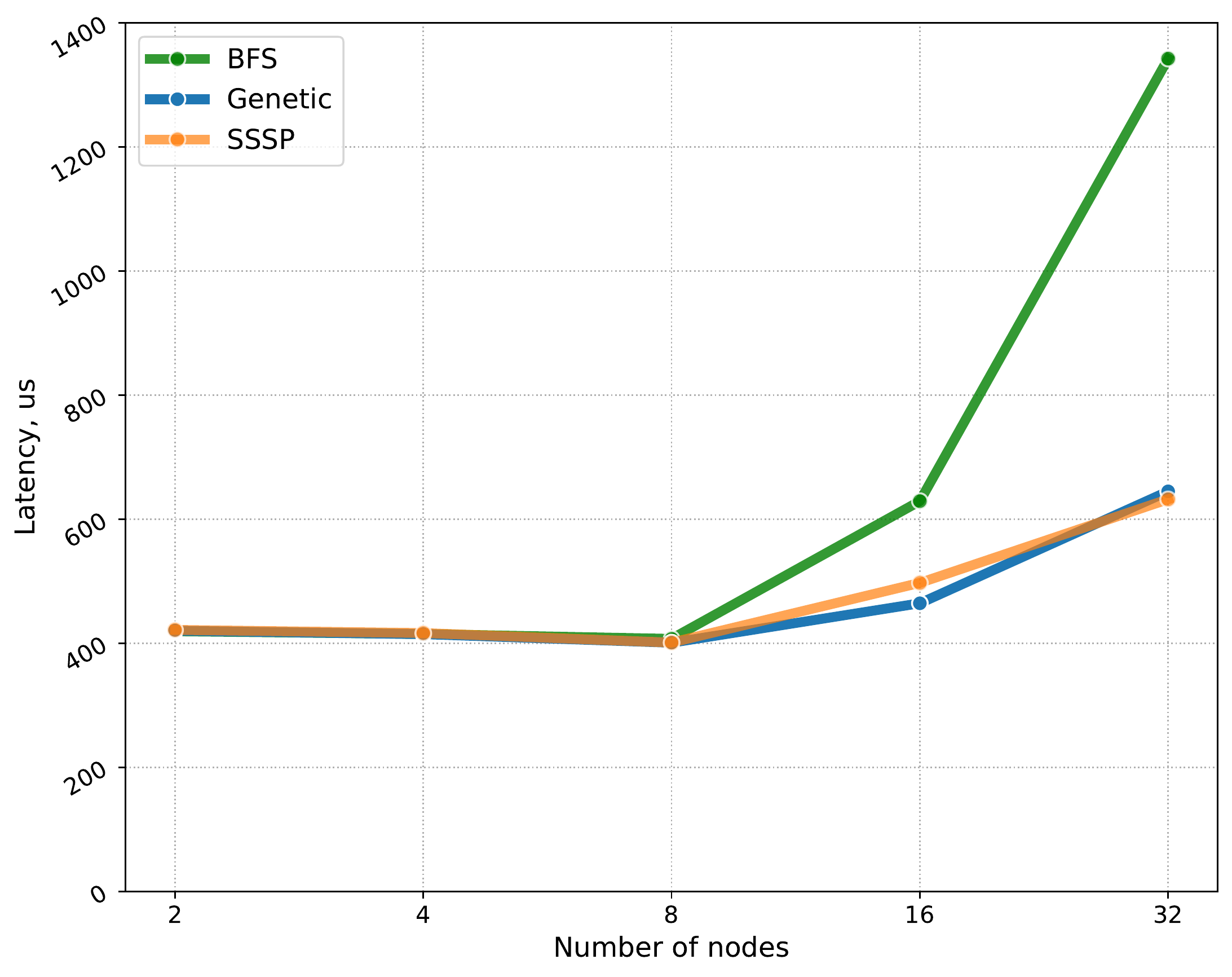}
		}
		\subfloat[Alltoall]{
			\label{fig:alltoall}
			\includegraphics[width=0.4\textwidth]{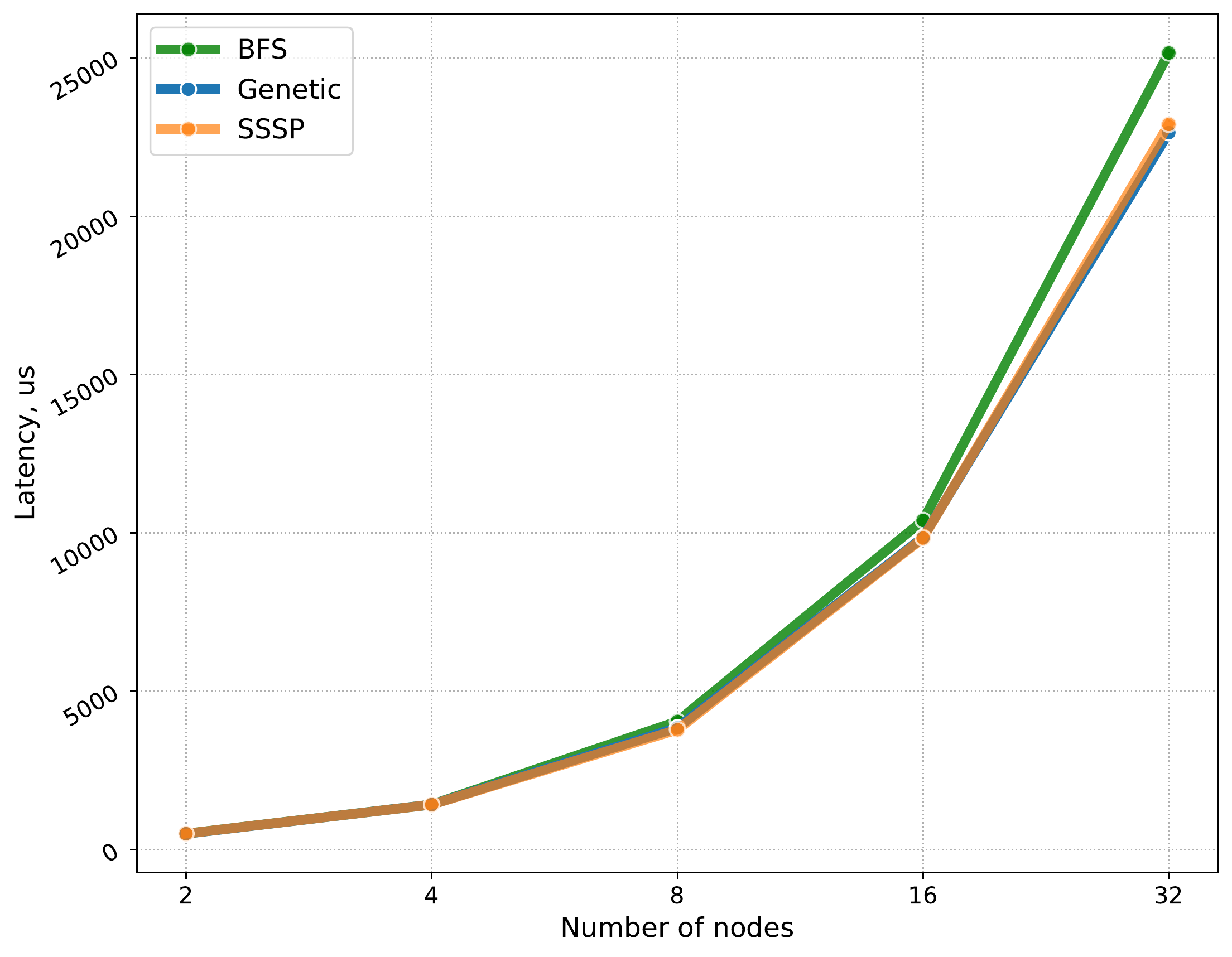}
		}
	\end{center}
	\caption{Comparison of the routing algorithms for communication patterns on the Desmos cluster.}
	\label{fig:patterns}
\end{figure*}

For the Desmos cluster we have chosen a node set for each number of nodes and provide to the SLURM the same node set for each benchmark. For each node number Table \ref{tab:Desmos_rt} compares the BFS, Genetic and SSSP algorithms. Maximum diameter $\max D$ of a routing table is the same for each node number, because the routing algorithms generate minimal paths. On 2 and 4 nodes all the algorithms obtain an optimal routing table, $max \gamma_c = min \gamma_c$ and $\sigma(4, R) = 0$. On 16 and 32 nodes the BFS algorithm is significantly worse than the Genetic and SSSP algorithms, the Genetic algorithm is slightly better than the SSSP algorithm.

To evaluate the routing table generation algorithms without application level impact we use four sythetic communication patterns: Transpose, Neighbor, Tornado and Alltoall \cite{dally2004principles}. The Transpose, Neighbor, Tornado patterns we implement using the MPI library. The Alltoall pattern implementation is an osu\_alltoall benchmark from OSU MicroBenchmarks version 5.6.2 \cite{osumb}. \FIXME{more comments, what are and why these patterns?}
\begin{table}[H]
	\centering
	\renewcommand{\arraystretch}{1.2}
	\caption{Comparison of routing table generation algorithms on the Desmos cluster.}
	\label{tab:Desmos_rt}
	\begin{tabular}{|m{1.2cm}|c|c|c|c|c|}
		\hline
		Routing Algorithm & \# nodes & $\max \gamma_c$ & $\min \gamma_c$ & $\sigma(4,\mathcal{R})$  & $\max D$ \\
		\hline
		\multirow{5}{*}{\textbf{BFS}}
		& 2  & 1  & 1 & 0 & 1 \\
		\cline{2-6}
		& 4  & 2  & 2 & 0 & 2 \\
		\cline{2-6}
		& 8  & 6  & 2 & 1.429 & 3 \\
		\cline{2-6}
		& 16 & 18 & 3 & 4.341 & 4 \\
		\cline{2-6}
		& 32 & 48 & 3 & 12.254 & 5 \\
		\hline
		\multirow{5}{*}{\textbf{Genetic}}
		& 2  & 1  & 1 & 0 & 1 \\
		\cline{2-6}
		& 4  & 2  & 2 & 0 & 2 \\
		\cline{2-6}
		& 8  & 4  & 4 & 0 & 3 \\
		\cline{2-6}
		& 16 & 11 & 5 & 1.745 & 4 \\
		\cline{2-6}
		& 32 & 27 & 8 & 6.274 & 5 \\
		\hline
		\multirow{5}{*}{\textbf{SSSP}}
		& 2  & 1  & 1 & 0 & 1 \\
		\cline{2-6}
		& 4  & 2  & 2 & 0 & 2 \\
		\cline{2-6}
		& 8  & 5  & 3 & 0.707 & 3 \\
		\cline{2-6}
		& 16 & 12 & 3 & 2.319 & 4 \\
		\cline{2-6}
		& 32 & 28 & 8 & 6.298 & 5 \\
		\hline
	\end{tabular}
\end{table}

\begin{table*}[t]
	\centering
	\renewcommand{\arraystretch}{1.2}
	\caption{NPB performance results on 32 nodes of the Desmos cluster for different routing table generation algorithms.}
	\label{tab:npb-results}
	\begin{tabular}{|c|c|c|c|c|c|c|c|}
		\hline
		& \multicolumn{2}{c|}{\textbf{BFS}} & \multicolumn{2}{c|}{\textbf{Genetic}} & \multicolumn{2}{c|}{\textbf{SSSP}} & Improvement in \%  of\\
		\cline{2-7}
		\multirow{1}{*}{\textbf{Benchmark}} & Mops & CoV & Mops & CoV & Mops & CoV & SSSP to BFS \\
		\hline
		LU & 364353 & 0.38\% & 367771 & 0.65\% & 368548 & 0.50\% & 101.15\% \\
		\hline
		MG & 374789 & 0.59\% & 376040 & 0.39\% & 377562 & 0.51\% & 100.74\% \\
		\hline
		CG & 88426 & 0.57\% & 90354 & 0.56\% & 90443 & 0.59\% & 102.28\% \\
		\hline
		FT & 172567 & 0.54\% & 181744 & 0.50\% & 181911 & 0.55\% & 105.41\% \\
		\hline
		IS & 7771 & 0.29\% & 8073 & 0.18\% & 8181 & 0.13\% & 105.28\% \\
		\hline
	\end{tabular}
\end{table*}

\newpage
Figure \ref{fig:patterns} compares the communication pattern results on the Desmos system. The message size is 1 MB, we perform 5 runs for each benchmark and node number. The running times are very stable, the coefficient of variation for a run does not exceed 0.71\%. On 2, 4 and 8 nodes the results are approximately the same. The Genetic algorithm is worse than BFS by 3.9\% on 32 nodes for the Transpose pattern (Figure \ref{fig:transpose}). For the other cases Genetic and SSSP are better than the BFS algorithm, on 32 nodes for the Neighbor pattern (Figure \ref{fig:neighbor}) the improvement is up to 11.6\%, for the Tornado pattern (Figure \ref{fig:tornado}) is up to 52.9\%, for the Alltoall pattern is 11.1\%. For the Alltoall pattern (Figure \ref{fig:alltoall}) Genetic outperforms the SSSP algorithm by 1.1\% on 32 nodes, the obtained routing table is slightly better balanced, which is characterized by $\max \gamma_c$ and $\sigma(4,\mathcal{R})$ values, see Table \ref{tab:Desmos_rt}. Coefficient of variation for that case is 0.08\% and 0.14\% for the Genetic and SSSP algorithms, respectively.

For the application benchmarks we use the MPI based
NAS Parallel Benchmarks suite \cite{npb} version 3.3.1, class C, 4 MPI processes per node. We use five parallel benchmarks LU, MG, FT, CG and IS, which represent wide different computation-to-communication ratio. We present in Table \ref{tab:npb-results} NPB performance results on 32 Desmos nodes and the relative performance improvement of the SSSP routing to BFS. Some benchmark results was unstable, for that reason we have adjusted a number of internal NPB iterations for each benchmark and have obtained fairly small coefficient of variation values. 

LU is a symmetric successive over-relaxation solver with a regular block matrix and exposes the high ratio between computation and communication. Observable performance improvement of SSSP can be explained by the benchmark sensitivity to small size message latency between neighbor nodes, recall the better SSSP routing table for the Neighbor pattern.

MG is a simplified multigrid method, it is a memory intensive benchmark. It requires highly structured short and long distance communication, but the whole volume of communication data is lower than for the LU benchmark. Therefore the performance difference for the considered routing algorithms is negligible.

CG is a conjugate gradient method with a large sparse symmetric positive-definite matrix. The communication pattern of CG is similar to MG, but the volume of communication data for CG is greater than for the MG benchmark \cite{riesen2006communication}. As a result, the SSSP based performance exceeds BFS based by 2.28\%.

FT solves a 3-dimensional partial differential equation using the fast Fourier transforms, IS is an integer sort. The FT and IS benchmarks use MPI collective operations mainly all-to-all communication. The advantage of the SSSP routing for the Alltoall synthetic benchmark leads to more than 5\% performance improvement for FT and IS on 32 nodes of the Desmos cluster.

\section{Conclusion}

We proposed the deadlock-free routing table generation
algorithm based on the fast SSSP algorithm for the deterministic routing of the torus topology Angara interconnect with a single virtual channel. The algorithm uses the routing graph that we proposed to provide a possibility to build routes in the torus topology Angara interconnect with a routing, that places restrictions on a further packet route with the previous route steps in mind. Compared to other BFS and Genetic routing algorithms in the most cases, on average, the edge forwarding index of the proposed algorithm is better than Genetic by 4\% and 7\% for 3D and 4D torus topologies, and is better than BFS by 13\%, 11\% and 17\% for 2D, 3D and 4D torus topologies.

We investigated the considered routing algorithms 
on the 32-node Desmos cluster system and benchmarked the proposed algorithm performance improvement of 11.1\% for the Alltoall communication pattern and of more than 5\% for the FT and IS application kernels.

The proposed routing table generation algorithm can be used in practice up to 150 nodes of 3D, 90 nodes of 4D and for 2D torus topology systems. In future works, we will try to improve a routing algorithm runtime to support large node number, preserving the routing quality.

The research was carried out with a grant from the Russian Science Foundation (project No. 20-71-10127).
\vspace{0.3cm}






	
	
	\bibliographystyle{IEEEtran}
	\bibliography{biblio}

\begin{thebibliography}{10}
\providecommand{\url}[1]{#1}
\csname url@samestyle\endcsname
\providecommand{\newblock}{\relax}
\providecommand{\bibinfo}[2]{#2}
\providecommand{\BIBentrySTDinterwordspacing}{\spaceskip=0pt\relax}
\providecommand{\BIBentryALTinterwordstretchfactor}{4}
\providecommand{\BIBentryALTinterwordspacing}{\spaceskip=\fontdimen2\font plus
\BIBentryALTinterwordstretchfactor\fontdimen3\font minus
  \fontdimen4\font\relax}
\providecommand{\BIBforeignlanguage}[2]{{%
\expandafter\ifx\csname l@#1\endcsname\relax
\typeout{** WARNING: IEEEtran.bst: No hyphenation pattern has been}%
\typeout{** loaded for the language `#1'. Using the pattern for}%
\typeout{** the default language instead.}%
\else
\language=\csname l@#1\endcsname
\fi
#2}}
\providecommand{\BIBdecl}{\relax}
\BIBdecl

\bibitem{simonov2014pervoye}
A.~Simonov, D.~Makagon, I.~Zhabin, A.~Shcherbak, E.~Syromyatnikov, and
  D.~Polyakov, ``Pervoye pokoleniye vysokoskorostnoy kommunikatsionnoy seti
  <<{A}ngara>>[the first generation of {A}ngara high-speed interconnect],''
  \emph{Naukoyemkiye tekhnologii [Science Technologies]}, vol.~15, no.~1, pp.
  21--28, 2014.

\bibitem{agarkov2016performance}
A.~Agarkov, T.~Ismagilov, D.~Makagon, A.~Semenov, and A.~Simonov, ``Performance
  evaluation of the {A}ngara interconnect,'' in \emph{Proc. Int. Conf. on
  Russian Supercomputing Days, Moscow, Russia}, 2016, pp. 626--639.

\bibitem{stegailov2019angara}
V.~Stegailov, E.~Dlinnova, T.~Ismagilov, M.~Khalilov, N.~Kondratyuk,
  D.~Makagon, A.~Semenov, A.~Simonov, G.~Smirnov, and A.~Timofeev, ``Angara
  interconnect makes {GPU}-based {D}esmos supercomputer an efficient tool for
  molecular dynamics calculations,'' \emph{The International Journal of High
  Performance Computing Applications}, 2019.

\bibitem{khalilov2018optimization}
M.~Khalilov and A.~Timofeev, ``Optimization of {MPI}-process mapping for
  clusters with {A}ngara interconnect,'' \emph{Lobachevskii Journal of
  Mathematics}, vol.~39, no.~9, pp. 1188--1198, 2018.

\bibitem{ostroumova2019reactive}
G.~Ostroumova, N.~Orekhov, and V.~Stegailov, ``Reactive molecular-dynamics
  study of onion-like carbon nanoparticle formation,'' \emph{Diamond and
  Related Materials}, vol.~94, pp. 14--20, 2019.

\bibitem{polyakov2018}
S.~Polyakov, V.~Podryga, and D.~Puzyrkov, ``High performance computing in
  multiscale problems of gas dynamics,'' \emph{Lobachevskii Journal of
  Mathematics}, vol.~39, no.~9, pp. 1239--1250, 2018.

\bibitem{stegailovvasp2019}
\BIBentryALTinterwordspacing
V.~Stegailov, G.~Smirnov, and V.~Vecher, ``{VASP} hits the memory wall:
  Processors efficiency comparison,'' \emph{Concurrency and Computation:
  Practice and Experience}, p. e5136, 2019. [Online]. Available:
  \url{https://doi.org/10.1002/cpe.5136}
\BIBentrySTDinterwordspacing

\bibitem{tolstykh2018structure}
M.~Tolstykh, G.~Goyman, R.~Fadeev, and V.~Shashkin, ``Structure and algorithms
  of slav atmosphere model parallel program complex,'' \emph{Lobachevskii
  Journal of Mathematics}, vol.~39, no.~4, pp. 587--595, 2018.

\bibitem{stegailov2020}
A.~Shamsutdinov, M.~Khalilov, T.~Ismagilov, A.~Piryugin, S.~Biryukov,
  V.~Stegailov, and A.~Timofeev, ``Performance of supercomputers based on
  angara interconnect and novel amd cpus/gpus,'' in \emph{International
  Conference on Mathematical Modeling and Supercomputer Technologies}.\hskip
  1em plus 0.5em minus 0.4em\relax Springer, 2020, pp. 401--416.

\bibitem{adiga2005blue}
N.~R. Adiga, M.~A. Blumrich, D.~Chen, P.~Coteus, A.~Gara, M.~E. Giampapa,
  P.~Heidelberger, S.~Singh, B.~D. Steinmacher-Burow, T.~Takken \emph{et~al.},
  ``Blue {G}ene/{L} torus interconnection network,'' \emph{IBM Journal of
  Research and Development}, vol.~49, no. 2.3, pp. 265--276, 2005.

\bibitem{scott1996cray}
S.~L. Scott \emph{et~al.}, ``The {C}ray {T3E} network: adaptive routing in a
  high performance 3{D} torus,'' 1996.

\bibitem{mukosey2019extended}
A.~Mukosey, A.~Simonov, and A.~Semenov, ``Extended routing table generation
  algorithm for the angara interconnect,'' in \emph{Russian Supercomputing
  Days}.\hskip 1em plus 0.5em minus 0.4em\relax Springer, 2019, pp. 573--583.

\bibitem{dally2004principles}
W.~J. Dally and B.~P. Towles, \emph{Principles and practices of interconnection
  networks}.\hskip 1em plus 0.5em minus 0.4em\relax Elsevier, 2004.

\bibitem{heydemann1989forwarding}
M.-C. Heydemann, J.~C. Meyer, and D.~Sotteau, ``On forwarding indices of
  networks,'' \emph{Discrete Applied Mathematics}, vol.~23, no.~2, pp.
  103--123, 1989.

\bibitem{sancho2003routing}
J.~C. Sancho, A.~Robles, P.~Lopez, J.~Flich, and J.~Duato, ``Routing in
  infiniband torus network topologies,'' in \emph{2003 International Conference
  on Parallel Processing, 2003. Proceedings.}\hskip 1em plus 0.5em minus
  0.4em\relax IEEE, 2003, pp. 509--518.

\bibitem{saad1993complexity}
R.~Saad, ``Complexity of the forwarding index problem,'' \emph{SIAM Journal on
  Discrete Mathematics}, vol.~6, no.~3, pp. 418--427, 1993.

\bibitem{montanana2009balanced}
J.~M. Monta{\~n}ana, M.~Koibuchi, H.~Matsutani, and H.~Amano, ``Balanced
  dimension-order routing for k-ary n-cubes,'' in \emph{2009 International
  Conference on Parallel Processing Workshops}.\hskip 1em plus 0.5em minus
  0.4em\relax IEEE, 2009, pp. 499--506.

\bibitem{kinsy2009application}
M.~A. Kinsy, M.~H. Cho, T.~Wen, E.~Suh, M.~Van~Dijk, and S.~Devadas,
  ``Application-aware deadlock-free oblivious routing,'' in \emph{Proceedings
  of the 36th annual international symposium on Computer architecture}, 2009,
  pp. 208--219.

\bibitem{abdel2011transcom}
A.~H. Abdel-Gawad and M.~S. Thottethodi, ``Transcom: Transforming stream
  communication for load balance and efficiency in networks-on-chip,'' in
  \emph{Proceedings of the 44th Annual IEEE/ACM International Symposium on
  Microarchitecture}, 2011, pp. 237--247.

\bibitem{abdel2016scalable}
A.~H. Abdel-Gawad, ``Scalable global optimal-bandwidth application-specific
  routing,'' in \emph{2016 IEEE 24th Annual Symposium on High-Performance
  Interconnects (HOTI)}.\hskip 1em plus 0.5em minus 0.4em\relax IEEE, 2016, pp.
  9--18.

\bibitem{dally1988deadlock}
W.~J. Dally and C.~L. Seitz, ``Deadlock-free message routing in multiprocessor
  interconnection networks,'' 1988.

\bibitem{schwiebert2001deadlock}
L.~Schwiebert, ``Deadlock-free oblivious wormhole routing with cyclic
  dependencies,'' \emph{IEEE Transactions on Computers}, vol.~50, no.~9, pp.
  865--876, 2001.

\bibitem{hoefler2009optimized}
T.~Hoefler, T.~Schneider, and A.~Lumsdaine, ``Optimized routing for large-scale
  infiniband networks,'' in \emph{2009 17th IEEE Symposium on High Performance
  Interconnects}.\hskip 1em plus 0.5em minus 0.4em\relax IEEE, 2009, pp.
  103--111.

\bibitem{dijkstra1959note}
E.~W. Dijkstra \emph{et~al.}, ``A note on two problems in connexion with
  graphs,'' \emph{Numerische mathematik}, vol.~1, no.~1, pp. 269--271, 1959.

\bibitem{domke2011deadlock}
\BIBentryALTinterwordspacing
J.~Domke, T.~Hoefler, and W.~E. Nagel, ``Deadlock-free oblivious routing for
  arbitrary topologies,'' in \emph{2011 {IEEE} International Parallel {\&}
  Distributed Processing Symposium}, IEEE.\hskip 1em plus 0.5em minus
  0.4em\relax {IEEE}, may 2011, pp. 616--627. [Online]. Available:
  \url{https://doi.org/10.1109/ipdps.2011.65}
\BIBentrySTDinterwordspacing

\bibitem{sancho2004effective}
J.~C. Sancho, A.~Robles, and J.~Duato, ``An effective methodology to improve
  the performance of the up*/down* routing algorithm,'' \emph{IEEE Transactions
  on Parallel and Distributed Systems}, vol.~15, no.~8, pp. 740--754, 2004.

\bibitem{sancho2000improving}
J.~C. Sancho and A.~Robles, ``Improving the up*/down* routing scheme for
  networks of workstations,'' in \emph{European Conference on Parallel
  Processing}.\hskip 1em plus 0.5em minus 0.4em\relax Springer, 2000, pp.
  882--889, \url{https://doi.org/10.1007/3-540-44520-X_123}.

\bibitem{dally1986torus}
W.~J. Dally and C.~L. Seitz, ``The torus routing chip,'' \emph{Distributed
  computing}, vol.~1, no.~4, pp. 187--196, 1986.

\bibitem{puente1999adaptive}
V.~Puente, R.~Beivide, J.~A. Gregorio, J.~Prellezo, J.~Duato, and C.~Izu,
  ``Adaptive bubble router: a design to improve performance in torus
  networks,'' in \emph{Proceedings of the 1999 International Conference on
  Parallel Processing}.\hskip 1em plus 0.5em minus 0.4em\relax IEEE, 1999, pp.
  58--67.

\bibitem{osumb}
OSU MicroBenchmarks. \url{http://mvapich.cse.ohio-state.edu/benchmarks/}.

\bibitem{npb}
D.~Bailey, T.~Harris, W.~Saphir, R.~Van Der~Wijngaart, A.~Woo, and M.~Yarrow,
  ``The nas parallel benchmarks 2.0,'' Technical Report NAS-95-020, NASA Ames
  Research Center, Tech. Rep., 1995.

\bibitem{riesen2006communication}
R.~Riesen, ``Communication patterns,''
  \url{http://citeseerx.ist.psu.edu/viewdoc/download?doi=10.1.1.420.5120&rep=rep1&type=pdf}.

\end{thebibliography}
	%
		
		

\end{document}